\documentclass[pdftex,twocolumn,prb,showpacs,preprintnumbers,amsmath,amssymb,superscriptaddress]{revtex4}

\pdfoutput=1
\usepackage{ifpdf}
\usepackage{graphicx}    
\usepackage{dcolumn}    
\usepackage{bm}             
\usepackage{url}
\usepackage{natbib}

\begin{document}

\title{Raman measurement of irreversible shear in SiO$_2$ glass}  

\author{Nikita S. Shcheblanov}
\affiliation{
Institut Lumi\`{e}re Mati\`{e}re, UMR 5306 Universit\'{e} Lyon 1-CNRS, \\
Universit\'{e} de Lyon F-69622 Villeurbanne Cedex, France
}
\author{Boris Mantisi}
\affiliation{
Laboratoire de Physique Th\'{e}orique de la Mati\`{e}re Condens\'{e}e, \\
Universit\'{e} Pierre et Marie Curie, 4 Place Jussieu, F-75252 Paris Cedex 05, France
}
\author{Paolo Umari}
\affiliation{
Dipartimento di Fisica e Astronomia, Universit\`{a} di Padova, via Marzolo 8, I-35131 Padova, Italy \\
and \\
CNR-IOM, Theory@Elettra group, Trieste, Italy
}
\author{Anne Tanguy}
\email{Anne.Tanguy@univ-lyon1.fr}
\affiliation{
Institut Lumi\`{e}re Mati\`{e}re, UMR 5306 Universit\'{e} Lyon 1-CNRS, \\
Universit\'{e} de Lyon F-69622 Villeurbanne Cedex, France
}

\date{\today}  
\begin{abstract}
Raman spectroscopy is a useful experimental tool to investigate local deformation and structural changes in SiO$_2$-based glasses. Using a semi-classical modelling of Raman spectra in large samples of silica glasses, we show in this paper that shear plastic flow affects the Raman measurement in the upper part of the spectrum. We relate these changes to structural modifications, as well as a detailed analysis of the vibration modes computed in the same frequency range. These results opens the door to in situ monitoring of plastic damage in silica-based structures. 
\end{abstract}

\keywords{silica glass, Raman spectra, defect lines, ring structures, molecular dynamics}

\pacs{62.20.F-, 78.30.-j, 78.30.Ly}
\maketitle

\section{INTRODUCTION}  

At present, amorphous materials continue to play a role of key materials in many technological areas, ranging from Si-based microelectronics to 5D Optical Data Storage by Ultrafast Laser Nanostructuring~\cite{zhang2014,gu2014}. Among them, silicate glasses are one of the most important, for example for their use as glass fibers, as dielectric insulator, or in the building industry. However these materials are brittle at large scale, with possibly catastrophic propagation of cracks, when they are submitted to large stresses. It is now well known that crack propagation in silicate glasses is initiated around plastic deformations occurring at the sub-micrometer scale~\cite{Hermansen2013}. The identification of plastic process zones would thus be of great interest in the effort for damage reduction in such materials. However, in situ measurements of plastic deformation are difficult at the micrometer scale. Thus, an heightened interest maintains nowadays to understand the structural atomic (re)arrangements in these materials. Diffraction probes like x-ray and neutron diffraction applied to amorphous materials can only partially characterize their atomic structure, since these materials lack translational symmetry. Therefore the vibrational spectroscopy techniques, in particular, Raman, Hyper-Raman, and Brillouin scatterings are applied as a complementary class of experimental techniques. Recent studies have revealed that Raman spectroscopy is not only sensitive to the local structure, but is a good local densification sensor in silicates~\cite{Champagnon2008}. In polymeric systems submitted to large shear deformation with shear banding, it was shown that Raman measurements are correlated to the direction of the shear band~\cite{Mermet1996}. In disordered materials, vibrational response is very sensitive to the proximity to plastic instability~\cite{tanguy2010}. However, shear deformation in silicate glasses takes place at a very small scale, and it is not clear wether the related structural changes would have a measurable Raman signature. Moreover, even with a visible signature of the deformation, it is rather difficult to extract directly structural informations from the experimental data, as this requires an accurate theoretical modelling of the material under investigation.

Our study is aimed to {investigate the microscopic basis of a signature of plastic shear }in SiO$_{2}$ glasses with the help of a modelling of Raman spectra coupled with classical molecular dynamics simulations. For crystalline solids, the first-principles methods have proved very successful in reproducing Raman cross-sections \cite{prb63}. However, the application of these approaches to calculate the Raman spectra of disordered materials has long been prevented by the computational effort required for treating systems of large system sizes. Furthermore, it was shown by one of us~\cite{anne2002} that long range correlations take place in the inhomogeneous vibrational response of glasses such that this size should be bigger than 50 \AA, that corresponds to $\sim$ 8000 atoms, to avoid the size-effects \cite{malavasi2006}. The paper is organized as follows: section 2 describes the preparation of the glass samples; section 3 gives details about semi-classical Raman simulations, and the calculation of the vibrational properties; in section 4 we show the results obtained on the calculation of Raman spectra, including a comparison with experimental results, and the sensitivity to plastic shear, with a description of the related vibration modes and structural changes. The last part serves to an additional discussion of the results in the light of previous modelling.

\section{Glass model} 

The amorphous silica model used in our numerical study was prepared following our previous work \cite{Boris2012}. An amorphous system of $N=8232$ atoms is obtained within classical molecular dynamics simulations using a melt-and-quench procedure. The model is contained in a box with linear size 50.16 \AA sufficiently large to get acoustic plane waves~\cite{anne2002}. The equilibration of the liquid, quench and relaxation of the glass are performed classically using the van Beest, Kramer and van Santen (BKS) potential \cite{BKS1990} modified by A. Carr\'{e} et al. \cite{Care2007}. It can be described as a two-body potential

\begin{align} \label {ex21}
  \Phi^{BKSW}_{\alpha  \beta}(r) = \Phi^{Coul}_{\alpha  \beta}(r) +  \Phi^{Buck}_{\alpha  \beta}(r),
\end{align}
where $\alpha$ and $\beta$ are the types of atoms (O or Si) and $r$ is the distance between them;

\begin{align} \label {ex221}
  \Phi^{Coul}_{\alpha  \beta}(r) = q_{\alpha}q_{\beta}e^{2}V_{W}(r)G_{W}(r),
\end{align}

\begin{align} \label {ex222}
  \Phi^{Buck}_{\alpha  \beta}(r)& = \left[ A_{\alpha  \beta}e^{-\frac{r}{\rho_{\alpha \beta}}} - \frac{C_{\alpha \beta}}{r^{6}}- \right.  \nonumber \\
  & \qquad  \left. - \left(A_{\alpha  \beta}e^{-\frac{r_{c,sh}}{\rho_{\alpha \beta}}} - \frac{C_{\alpha \beta}}{r^{6}_{c,sh}} \right) \right]G_{sh}(r),
\end{align}
with

\begin{gather} \label {ex23}
 V_{W}(r) = \left(\frac{1}{r} - \frac{1}{r_{c,W}} \right) + \frac{1}{r^{2}_{c,W}} ( r - r_{c,W}), \\
 G_{W}(r) = exp \left(-\frac{\gamma^{2}_{W}}{( r - r_{c,W})^{2}} \right),\\
 G_{sh}(r) = exp \left(-\frac{\gamma^{2}_{sh}}{( r - r_{c,sh})^{2}} \right),
\end{gather}
where $\gamma_{sh} = \gamma_{W} = 0.5$, $r_{c,W} = 10.17$ \AA, $r_{c,sh} = 5.5$ \AA. 

We also add a strong and regular repulsive part at short range $(r < r_{inf})$ to avoid the collapse of atoms at high pressure, or high temperatures. The added repulsive part has the following form:

\begin{equation} \label {ex24}
 \Phi^{Rep}_{\alpha  \beta}(r) = \left ( \frac{D_{\alpha  \beta}}{r} \right)^{12}+E_{\alpha \beta}r+F_{\alpha  \beta}.
\end{equation}
$D_{\alpha  \beta}$, $E_{\alpha  \beta}$, and $F_{\alpha  \beta}$ have been adjusted in order to have the continuity of the potential and its first, and second derivatives. The parameters of this potential are tabulated in Table~\ref{Tab1}.

The total energy can then be written

\begin{equation}
E_{tot}=\sum_{i<j} \Phi^{BKSW}_{\alpha_i \beta_j}(r_{ij})+\Phi^{Rep}_{\alpha_i \beta_j}(r_{ij})
\end{equation}
and will be used to compute the Dynamical Matrix, as will be explained later.

\begin{table} [!ht]
\resizebox{8.5cm}{!} {
\begin{tabular}{ccccc}
\toprule
\\
 $   $& $A_{\alpha \beta}$ (eV) & $\rho_{\alpha \beta}$ (\AA)  & $C_{\alpha \beta}$ (eV \AA$^{6}$) & $D_{\alpha \beta}$ (\AA eV$^{-12}$)  \\
\hline
\\
O-O & 1388.773 & 0.3623 & 175.0 & 1.51166281 \\
Si-O & 18003.7572 & 0.2052 & 133.5381 & 1.42402882 \\
Si-Si & 872360308.1 & 0.0657 &23.299907 & 0.0 \\
\\
\hline
\\
 $   $& $E_{\alpha \beta}$ (eV \AA$^{-1}$) & $F_{\alpha \beta}$ (eV)  & $r_{inf}$ (\AA) &   \\ 
\hline
\\
O-O & -14.97811134 & 39.0602602165 & 1.75 & \\
Si-O & -3.24749265 & -15.86902056 & 1.27 & \\
Si-Si & 0.0 & 0.0 & 0.0 & \\ 
\\
\toprule
\end{tabular}
}
\caption{\label{Tab1}Parameters of the empirical potential used to model the silica glass.}
\end{table}

\begin{figure}
\begin{center}
\includegraphics[width=8.5cm] {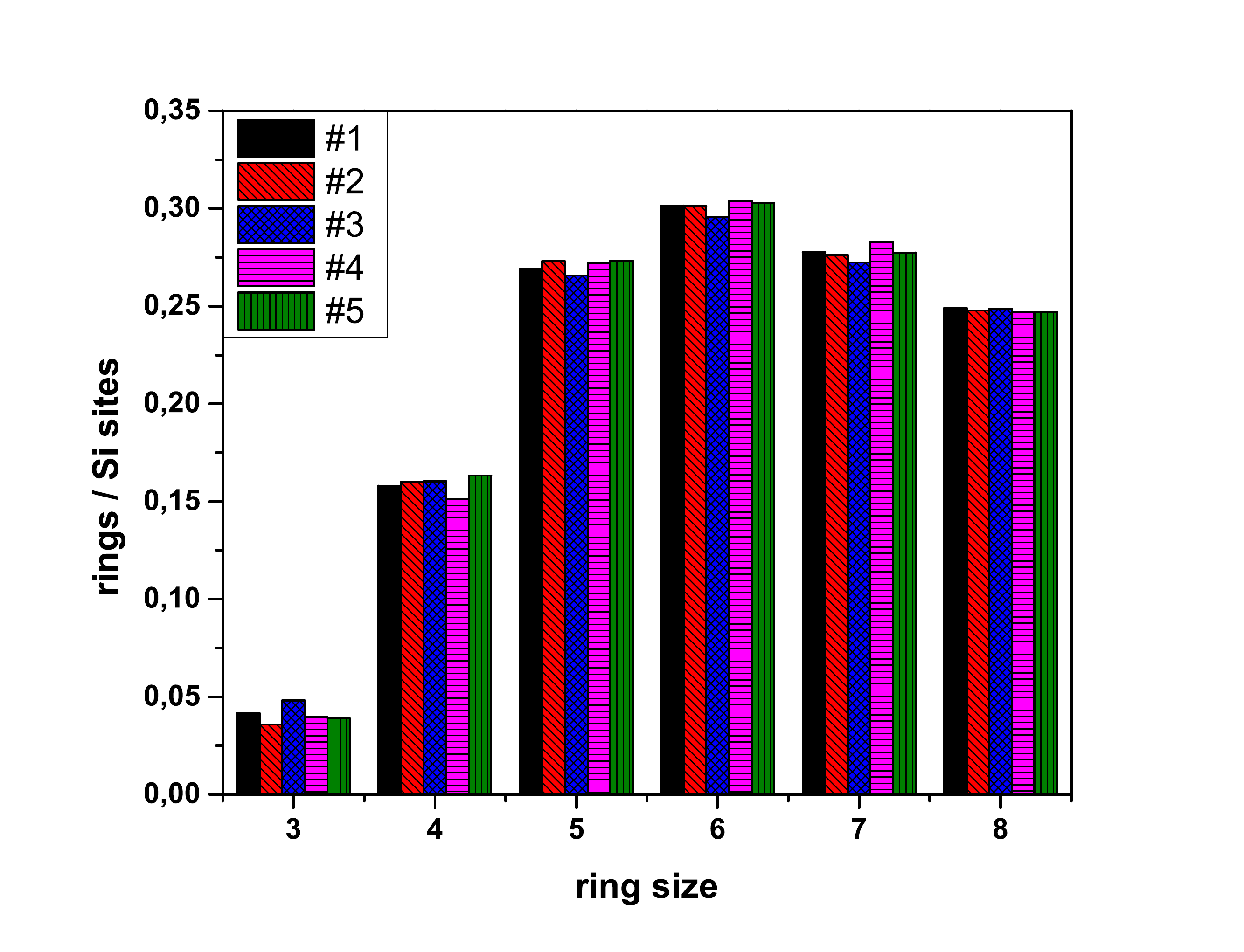} 
\caption{\label{figrings-basic} Ring statistics for uncompressed a-SiO$_{2}$ sample (8232 particles).}
\end{center}
\end{figure}

\begin{figure}
\begin{center}
\includegraphics[width=8.5cm]{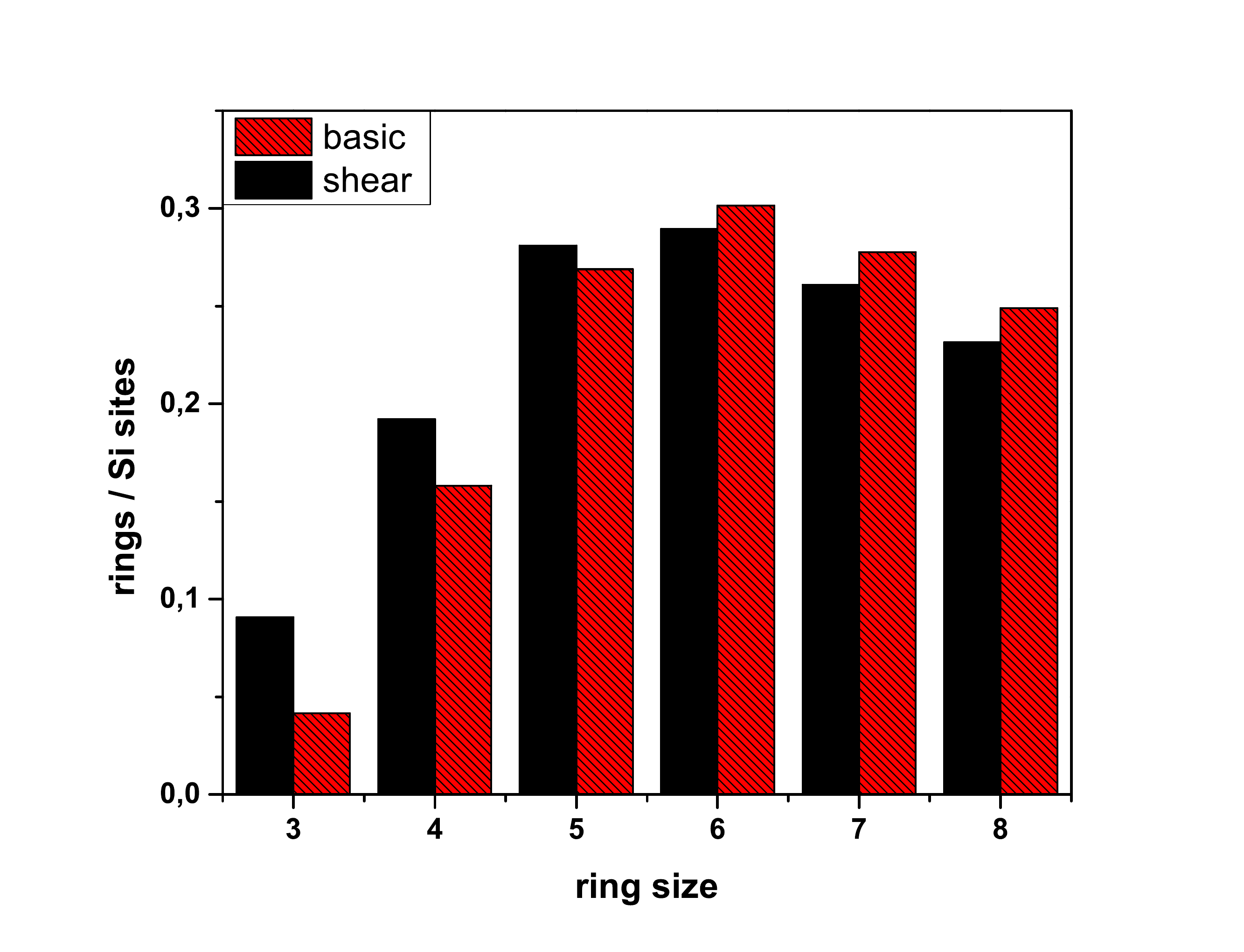}
\caption{\label{figrings-shear}Rings statistics in a-SiO$_{2}$ upon shear. Comparison between the initial distribution of rings (red) and the distribution of rings after the sample entered in the plastic plateau (black).}
\end{center}
\end{figure}

\begin{figure*}
\begin{center}
\includegraphics[width=16cm]{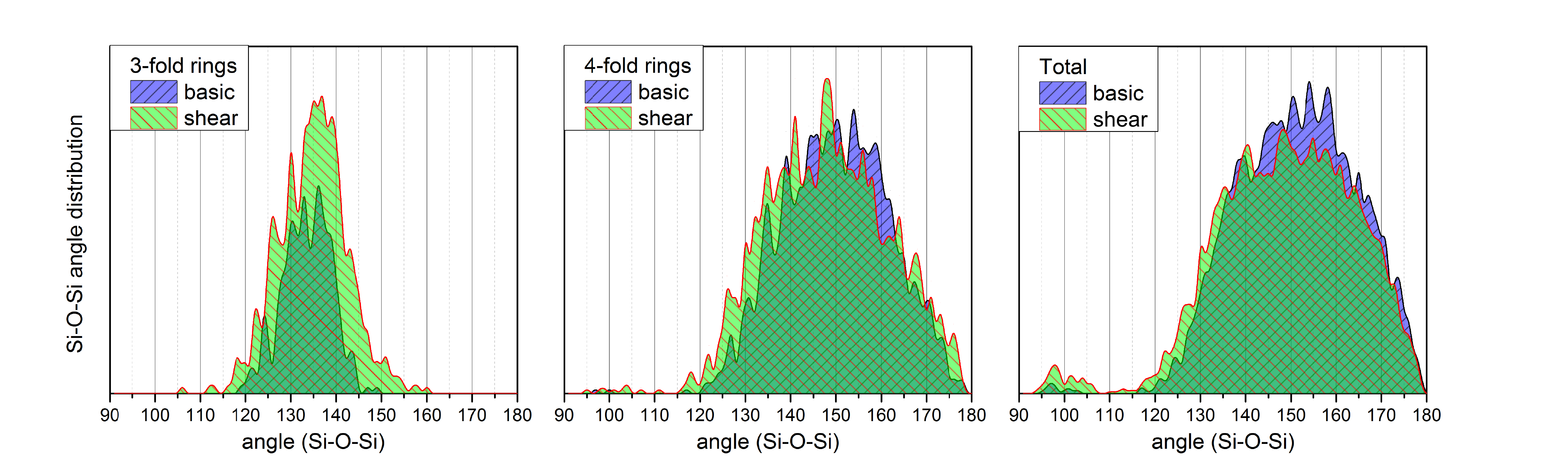}
\caption{\label{fig.angles} Distribution of Si-O-Si angles (a) in 3-fold rings, (b) in 4-fold rings, and (c) total distribution.}
\end{center}
\end{figure*}

\begin{figure}
\begin{center}
\includegraphics[width=8.5cm]{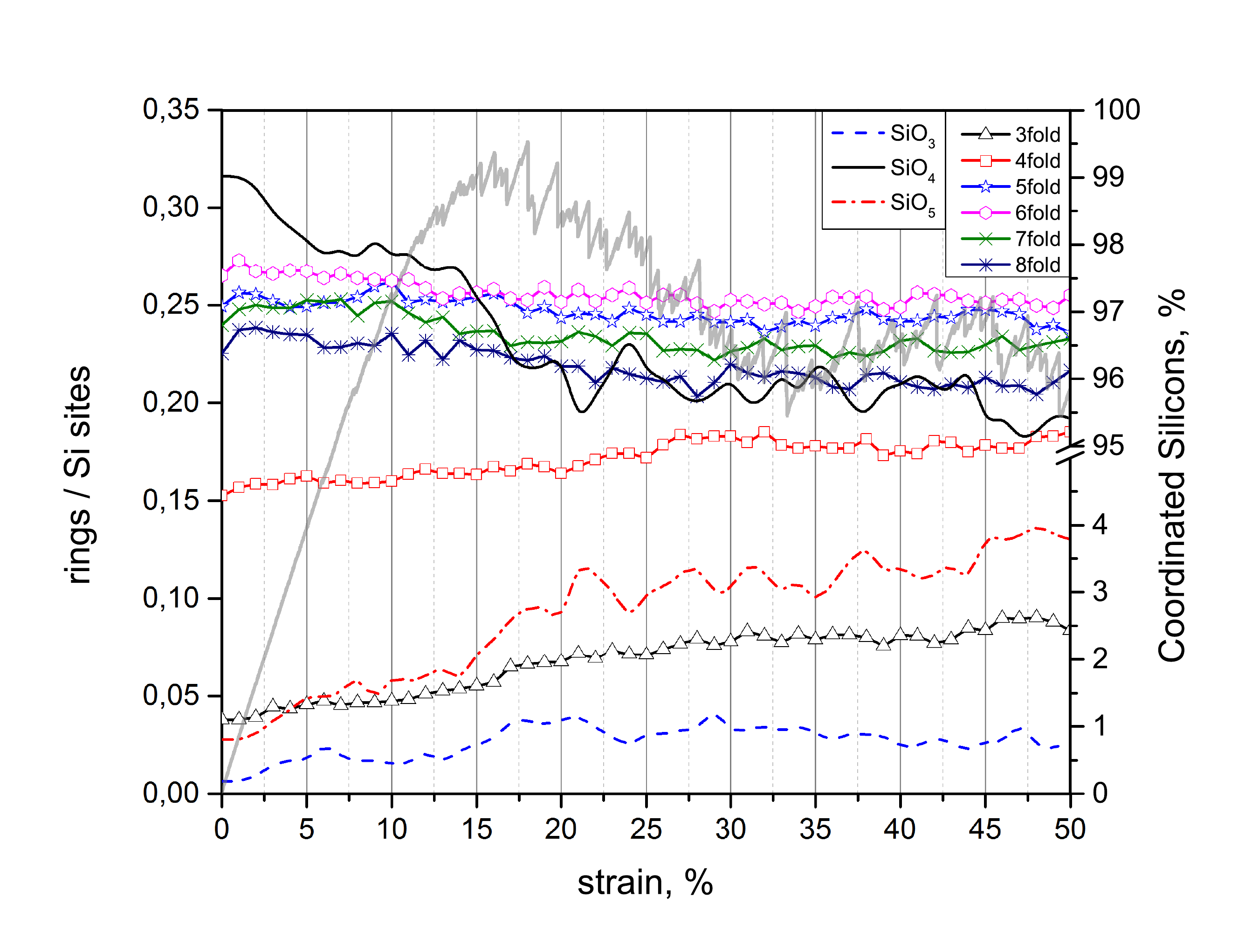}
\caption{\label{fig-Coord-Rings} Evolution of the coordination number, and rings distribution in a-SiO$_2$ model sample, upon athermal plastic shear deformation.}
\end{center}
\end{figure}

The sample preparation proceeds as follows. The glasses are obtained from a crystal state which is heated and then quenched. We start from a $\beta$-Cristobalite crystal sample, which is the last crystalline state before the melting point at normal pressure. Then we heat the sample up to 5200 K during 1 ns, let it evolve at constant temperature during 10 000 time steps (time step $\delta t = 10^{-15}$ s), and quench it at 0 K in 1 ns. The quenching rate is hence 5.2$\times 10^{12}$ K s$^{-1}$. Finally we relax the simulation box to avoid residual stress, and then obtain a density of 2.18 g cm$^{-3}$ with a pressure about 0 GPa with a precision of 10 MPa. The structural properties of such system have already been studied extensively~\cite{Boris2012}. We show in Fig.~\ref{figrings-basic} the distribution of {\it Si}-rings in the system. The {\it Si}-rings describe the medium-range order~\cite{luigi2009} of vitreous SiO$_{2}$. They are defined as the smallest closed loop obtained starting on a Si atom and following Si-O bonds~\cite{ring2010}. A 3-membered (or 3-fold) ring contains three Si atoms, a n-membered (n-fold) ring contains n Si atoms. The distribution of rings in silica glasses has often been related to the amplitude of $D1$ and $D2$ bands in Raman spectra~\cite{galeener85,alf1998,alf2002,umari2003}. In Fig.~\ref{figrings-basic}, we see that for the sample studied at rest in this paper, the distribution is maximum for 6-membered rings, and the proportion of 5- and 7- membered rings is approximately the same. This distribution is sensitive to the pressure, the density, and the quenching rate used during the sample's preparation. We see in this paper that it depends as well on the plastic shear (Fig.~\ref{figrings-shear}), with an increasing number of small rings upon shear. It should be noted, that previously \cite{smirnov2008}, the glass-samples prepared for a number of particles smaller than 100, had significant variations in rings distribution conditionally on the time spent in the liquid state. However, it was mentioned that this distribution depends on the number of particles \cite{smirnov2008}. We found out that the variation becomes weak (less sensitive to the time spent in the liquid state during the preparation of the glass (see Fig.~\ref{figrings-basic}) with an increasing number of particles. The obtained distribution is in agreement with the last results of D.~Waroquiers~\cite{david2013}, and a similar distribution was also obtained in the recent experimental results of monolayer SiO$_{2}$~\cite{huang2012,wilson2013,buchner2014}. For unknown reasons, the number of small 3- and 4-membered rings is always larger in our systems than in experimental measurements. We will comment later it's evolution upon plastic shear deformation. The distribution of Si-O-Si angles is shown in Fig.~\ref{fig.angles}. The total distribution is in agreement with previous results~\cite{Boris2012} with a maximum at $150^\circ\pm 20^\circ$, and agrees reasonably well with experimental results~\cite{coombs2008,mozzi1969}. The distribution of Si-O-Si angles in 3-fold rings is more peaked and centered on lower values ($135^\circ\pm5^\circ$), while the distribution of Si-O-Si angles in 4-fold rings is very close to the total one. Note that the values obtained in our samples for the distributions restricted to 3-fold and 4-fold rings are larger that those obtained in the preceding work of A. Pasquarello et al.~\cite{alf1998} and closer to the experimental one. Upon shear, the distribution shows a secondary distribution centered on $100^\circ$. This auxiliary angle corresponds to adjacent tetrahedra, and was already seen upon compression~\cite{Boris2012} but in larger proportions. Upon shear, the effect is only slightly visible, and affects mainly Si-O-Si angles belonging to rings whose size is larger than 4-fold. In the same time, it was shown (see Fig.~\ref{fig-Coord-Rings}) that only a few percent of particles vary its coordination number: $3\%$ additional 5-coordinated Si-atoms (as discussed in~\cite{huang2012}) appear upon shear in the sample. Local structural changes are thus only tiny visible upon shear at constant volume. We will discuss later the more pronounced evolution of mesoscopic structural correlations (evolution in the number of rings) upon shear.

We have chosen to perform the mechanical deformation of our systems in the a-thermal regime, i.e. we either perform energy minimization (T = 0 K), or we let our glass evolve at a very low temperature T $= 10^{-5}$ K, temperature at which thermal effects are totally negligible and not high enough to activate plastic events~\cite{Tanguy2011}. For the shear at constant volume, we changed the shape of the box within a monoclinic symmetry and a small deformation step, and then let the atoms relax by minimizing the total energy in the box at constant volume. We used the Polak-Ribiere version of the conjugate gradient (CG) minimization algorithm proposed by LAMMPS. {At each iteration, the force gradient is combined with the previous iteration information to compute a new search direction perpendicular (conjugate) to the previous search direction.} We carry on this operation to get the desired total shear deformation. The shear step corresponds to a deformation of $\Delta \epsilon_{xy} = 10^{-4}$, that is small enough to decorrelate the plastic events that occur during the shear \cite{Tanguy2005}. It corresponds to a very fast relaxation time between each strain step (quasi-static deformation) that allows to follow a mechanical equilibrium path (athermal response). This kind of deformation gives very detailed results on the restructuring of the phase space upon mechanical deformation \cite{Tanguy2011}, and allows to construct the landscape over which thermal activation acts. Thermal activation, that can play a role even below the glass transition temperature \cite{Tanguy2011}, will thus not be discussed here. 

\begin{figure} 
\begin{center}
\includegraphics[width=8.5cm]{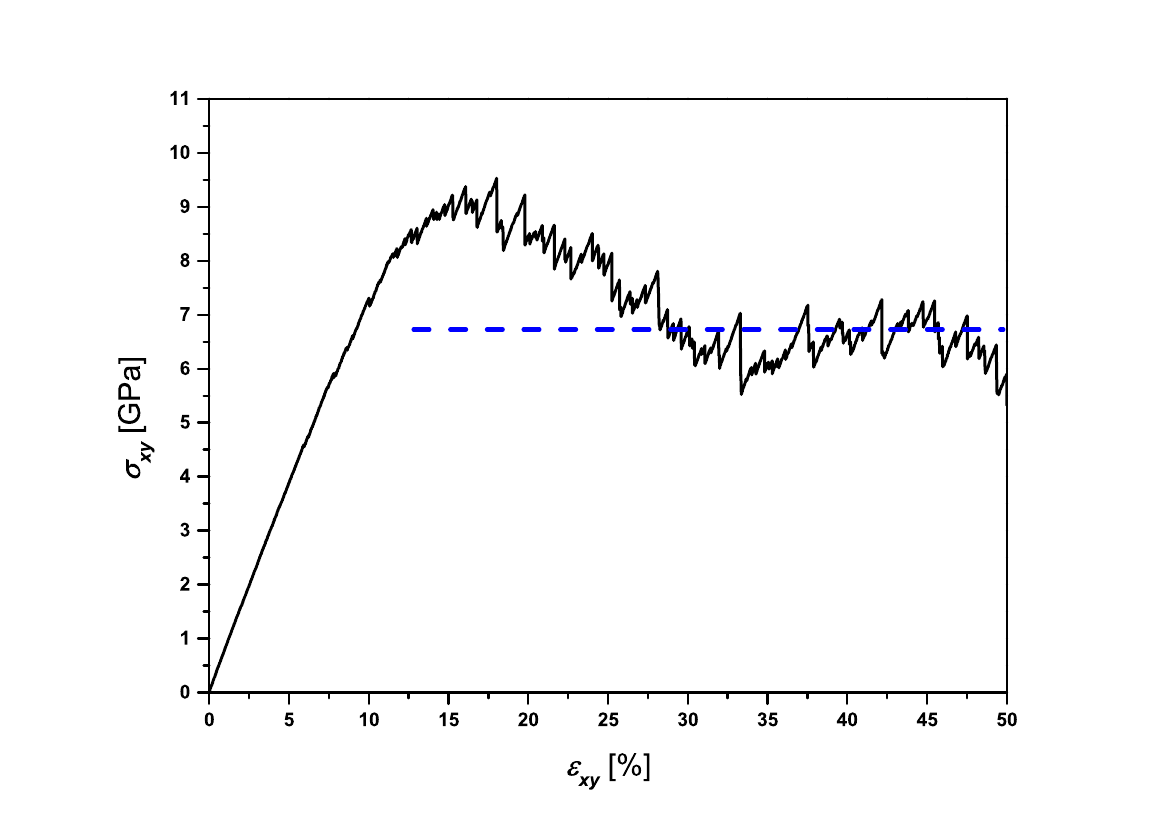}
\caption{\label{figstress} Shear stress as a function of engineering shear strain in a-SiO$_{2}$ sample upon quasi-static and athermal shear deformation.}
\end{center}
\end{figure}

The resulting stress-strain curve is shown in Fig.~\ref{figstress}. It shows a linear response followed by stress-softening and hatched stress relaxation processes. At large deformation ($>30\%$) the system reaches a visco-plastic plateau characterized by a constant average visco-plastic flow stress. The hatches, that are the fluctuations around the average flow, correspond to local plastic rearrangements giving rise to energy dissipation~\cite{Boris2012}. We compared the results obtained on 10 initial configurations at rest, and 10 configurations chosen randomly in the plastic plateau. 

\section{Simulation details} 
\subsection {Raman model}
\label{part:RamanModel}

Here we briefly outline the formulation that we used for the calculation of Raman activities. We only focus on first-order processes, which involve a single phonon excitation. For a model system, the vibrational frequencies $\omega_n$ and their associated eigenmodes $\xi^{n}_{\alpha}$ are found by solving the set of linear equations:

\begin{equation} \label {ex1}
\sum_{B j} D_{A i,B j} \xi^{n}_{B j} = \omega^{2}_{n}\xi^{n}_{A i},
\end{equation}
where $\bf {D}$ is the dynamical matrix, and where the capital Latin indices run over the atoms, and the uppercase Latin indices are the three Cartesian directions, respectively. The dynamical matrix $\bf {D}$ is defined by

\begin{equation} \label {ex2}
D_{A i,B j} = \frac{1}{\sqrt{M_{A}M_{B}}} \frac{\partial^2E_{tot}}{\partial R_{A i}\partial R_{B j}},
\end{equation}
where $E_{tot}$ is the total energy of the system, and ${\bf R_A}$ and $M_A$ are the position and the atomic mass of atom $A$. Diagonalization of the dynamical matrix $\bf {D}$ gives the vibrational frequencies $\omega^{2}_{n}$ and the normalized eigenmodes $\xi^{n}_{A i}$. The Raman susceptibility tensors \cite{cardona1982} $\mathcal{R}^n_{ij}$ are then given for each mode $n$ by

\begin{equation} \label {ex3}
\mathcal{R}^n_{ij} = {\sqrt{V}}\sum_{A,k} {\frac{\partial \chi_{ij}}{\partial R_{A k}}\frac{\xi^{n}_{A k}}{\sqrt{M_A}}},
\end{equation}
where $\chi$ is the electric polarizability tensor and $V$ is the volume of the cell.

In experimental set-ups, it is customary to record the Raman spectra in the HH and the HV configurations, in which the polarization of the outgoing photons is respectively parallel and orthogonal to the ingoing photon polarization~\cite{bruesch1986}. Using the isotropy of disordered solids, we express the contributions of the $n$-th mode to 
the HH and HV Raman spectra as \cite{cardona1982}

\begin{gather} \label {ex4}
I^{n}_{HH} = a^{2}_{n} + \frac{4}{45}b^{2}_{n}, \\
I^{n}_{HV} = \frac{3}{45}b^{2}_{n},
\end{gather}
where $a_n$ and $b_n$ are obtained from

\begin{equation} \label {ex6}
a_{n} = \frac {1}{3}\left(\mathcal{R}^{n}_{11}+\mathcal{R}^{n}_{22}+\mathcal{R}^{n}_{33}\right), \\
\end{equation}

\begin{multline}
b^{2}_{n} = \frac {1}{2}\left[ \left(\mathcal{R}^{n}_{11}-\mathcal{R}^{n}_{22}\right)^2+\left(\mathcal{R}^{n}_{11}-\mathcal{R}^{n}_{33}\right)^2+ \right. \\
\left. +\left(\mathcal{R}^{n}_{22}-\mathcal{R}^{n}_{33}\right)^2\right]+3\left[\left(\mathcal{R}^{n}_{12}\right)^2+\left(\mathcal{R}^{n}_{13}\right)^2+\left(\mathcal{R}^{n}_{23}\right)^2 \right]
\end{multline}

Focusing on the Stokes process, in which a vibrational excitation is created by an incoming photon, we express the total power cross-section as (in esu units)

\begin{multline} \label {ex8}
I^{P}_{HH,HV}(\omega) = 4\pi \sum_{n} \frac{(\omega_{L}-\omega_{n})^4V}{c^4}\times \\
\times I^{k}_{HH,HV}\frac{\hbar}{2\omega_n}[n(\omega_n)+1]\delta(\omega-\omega_n),
\end{multline}
where $V$ is the volume of the scattering sample, $\omega_L$ the frequency of the incoming photon, $c$ the speed of light, and $n(\omega)$ the boson factor:
\begin{equation} \label {ex9}
n(\omega) = \frac{1}{exp\left(\hbar\omega/k_{B}T\right)-1}.
\end{equation}

The bond polarizability model \cite{wolk1941,wolk1944} has successfully been applied for the calculation of Raman intensities in a large variety of systems \cite{cardona1982}. In this approach, 
the polarizability is modeled in terms of bond contributions:

\begin{equation} \label {ex10} 
\chi_{ij}(A) = \frac{1}{V}\sum_B\alpha_{ij}(A,B)
\end{equation}
with

\begin{multline}
\alpha_{ij}(A,B)=\frac{1}{3}(2\alpha_p+\alpha_l)\delta_{ij} +\\
+ (\alpha_l-\alpha_p)\left(\frac{R_iR_j}{{\left | {\bf R} \right |}^2}-\frac{1}{3}\delta_{ij}\right),
\end{multline}
where $\bf {R=R_A-R_B}$ is a vector which defines the direction and the distance of a pair of nearest neighbor atoms at sites $\bf {R_A}$ and $\bf {R_B}$. The parameters $\alpha_l$ and $\alpha_p$ correspond to the longitudinal and perpendicular bond polarizability, respectively.

The bond polarizability model further assumes that the bond polarizabilities $\alpha_l$ and $\alpha_p$ only depend on the length of the bond. Thus, the derivative of the local bond polarizability with respect to the relative displacement of the atoms $A$ and $B$ yields:

\begin{multline} \label {ex11}
\frac{\partial \alpha_{ij}(A,B)}{\partial R_{A k}} = \frac{1}{3}\left(2\alpha'_{p}+\alpha'_{l}\right)\delta_{ij}\hat{R_k} +\\
+\left(\alpha'_l-\alpha'_p\right)\left(\hat{R_i}\hat{R_j}-\frac{1}{3}\delta_{ij}\right)\hat{R_k} +\\
+\frac{(\alpha_l-\alpha_p)}{R}\left(\delta_{ik}\hat{R_j}+\delta_{jk}\hat{R_i}-2\hat{R_i}\hat{R_j}\hat{R_k}\right),
\end{multline}
where $\bf{\hat{R}}$ is a unit vector along $\bf {R=R_A-R_B}$, and $\alpha'_{l,p}$ are the derivatives of the bond polarizabilities with respect to the bond length ($\alpha'_{l,p} = (\partial \alpha_{l,p} / \partial R ) |_{R=R_{0}} $, where $R_{0}$ is a typical distance). Therefore, when one type of bond occurs, the bond polarizability model is completely defined by three parameters: $2\alpha'_p + \alpha'_l$, $\alpha'_l - \alpha'_p$, and  $(\alpha_l - \alpha_p) / R$. We used the parameters of the bond polarizability model already derived by one of us in Ref. \cite {luigi2007,paolo2003}, whose values are summarized in Table~\ref{Tab2}.

\begin{table} [!ht]
\caption{\label{Tab2}Bond-polarizability-model parameters.}
\setlength{\tabcolsep}{8pt}
\begin{tabular}{ c c c c }
\toprule
\\
Model & $2\alpha'_{p} + \alpha'_l$ & $\alpha'_{l} - \alpha'_{p}$ & $(\alpha_{l} - \alpha_{p}) / R$  \\ 
\\
\hline \\
BP & 0.771 & 0.196 & 0.056 \\ 
($(4\pi)^{-1} \cdot $Bohr$^{-1}$) & & & \\
\\
\toprule
\end{tabular}
\end{table}

\subsection{Vibrational density of state}

The frequencies $\omega_n$ and the corresponding normalized eigenmodes ${\bf \xi}^{n}_{\bf A}$ are obtained by diagonalizing the dynamical matrix. The Intel MKL tool was used for the diagonalization \cite{intel}. The associated atomic displacements are given by:

\begin{equation} \label {ex13}
{\bf u}^{n}_{\bf A} = \frac{{\bf \xi}^{n}_{\bf A}}{\sqrt{M_A}}.
\end{equation}
The index $n$ labeling the vibrational modes runs from 1 to $3N$, where $N$ is the total number of atoms in the model.

The $v$-DOS can be decomposed according to the weights of the two species: $Z(\omega) = \sum_\alpha{Z_\alpha(\omega)}$. The partial density of states $Z_\alpha(\omega)$ is defined by:

\begin{equation}
Z_\alpha(\omega) = \frac {1}{3N}\sum^{N_\alpha}_{A}{\sum^{3N}_{n} {{\left | {\bf \xi}^{n}_{\bf A} \right|}^2\delta \left(\omega-\omega_n\right)}},
\label {ex22}
\end{equation}
where $A$ runs over the set of atoms belonging to a given species $\alpha$. The $v$-DOS can be decomposed as well according to the weight of different kinds of vibrations (e.g. bending, rocking, stretching). The vibrations of the bridging oxygens (BO) may be classified as stretching, bending, or rocking according to whether the BO atoms move parallel to the line joining the the two Si neighbors, parallel to bisector of the Si-O-Si angle or perpendicular to the Si-O-Si plane, respectively. Thus, if we apply following unit vectors: $\bf{\hat{r}_{\oplus}}$ the unit vector perpendicular to the local Si-O-Si plane, $\bf{\hat{r}_{i+i'}}$, the unit vector parallel to the bisector of the Si-O-Si angle, and $\bf{\hat{r}_{i-i'}}$ the unit vector perpendicular to  $\bf{\hat{r}_{\oplus}}$ and $\bf{\hat{r}_{i+i'}}$, then the corresponding projection of the vibration modes in the rocking, stretching and bending and contributions contribute to the partial density of states as

\begin{equation}
Z_{O}(\omega) = Z_{rock}(\omega)+Z_{str}(\omega)+Z_{ben}(\omega)
\end{equation}
with $\alpha$ retricted to Oxygen atoms, and

\begin{eqnarray}
Z_{rock}(\omega)= \frac {1}{3N}\sum^{N_O}_{A=Ox}{\sum^{3N}_{n} {{\left | {\bf \xi}^{n}_{\bf A} \cdot \bf{\hat{r}_{\oplus}} \right|}^2\delta \left(\omega-\omega_n\right)}}, \label{eq-R} \\
Z_{str}(\omega)=\frac {1}{3N}\sum^{N_O}_{A=Ox}{\sum^{3N}_{n} {{\left | {\bf \xi}^{n}_{\bf A} \cdot \bf{\hat{r}_{i-i'}} \right|}^2\delta \left(\omega-\omega_n\right)}}, \label{eq-S} \\
Z_{ben}(\omega)=\frac {1}{3N}\sum^{N_O}_{A=Ox}{\sum^{3N}_{n} {{\left | {\bf \xi}^{n}_{\bf A} \cdot \bf{\hat{r}_{i+i'}} \right|}^2\delta \left(\omega-\omega_n\right)}}. \label{eq-B}
\end{eqnarray}

\section{RESULTS} 

\subsection{Validity of the semi-classical calculation}

\begin{figure} 
\begin{center}
\includegraphics[width=8.5cm]{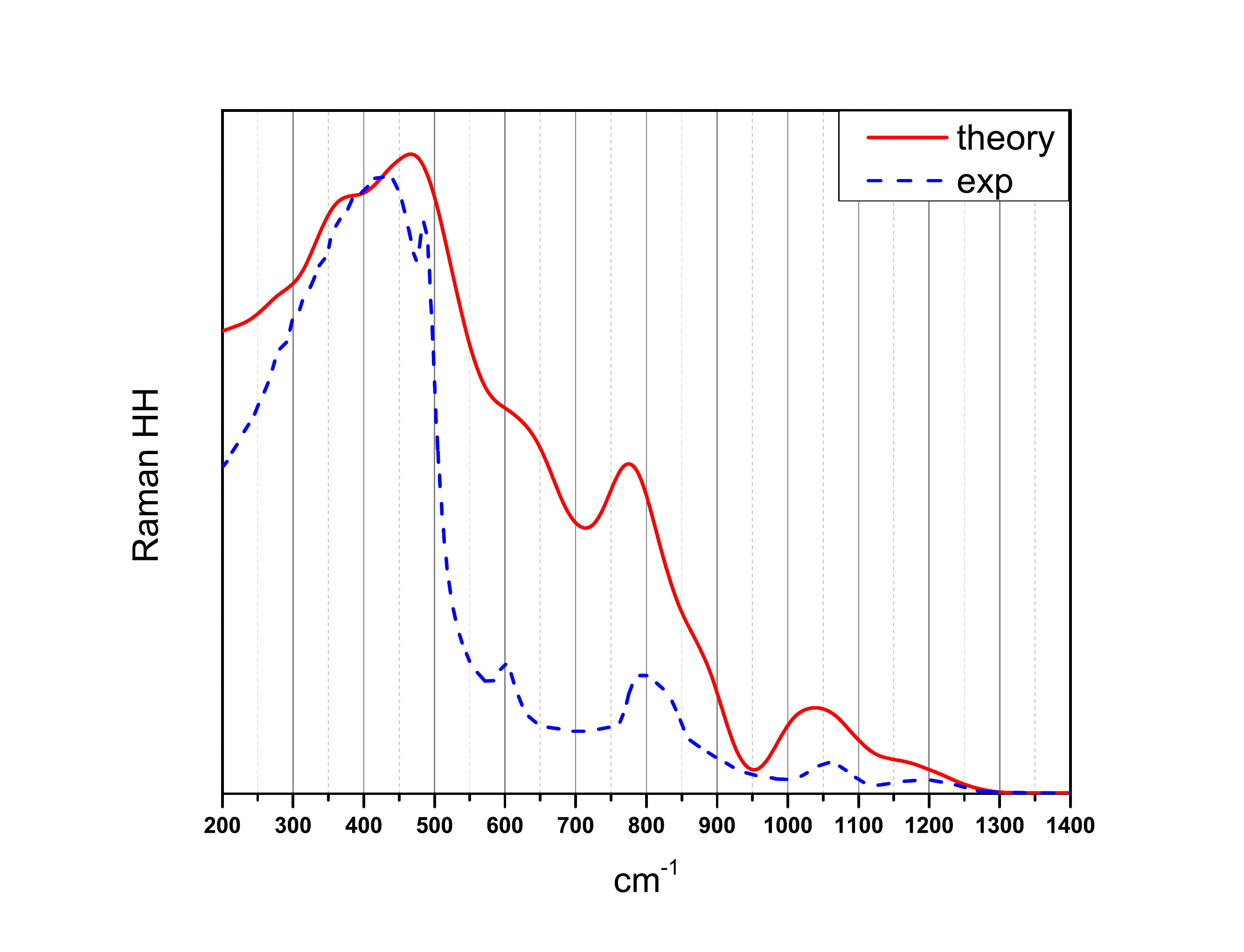} 
\caption{\label{figHHexp} Raman HH spectrum in uncompressed a-SiO$_{2}$. Red (solid) curve is theory [current work], black (dash) curve is experimental data from~\cite{sonneville2013}.}
\end{center}
\end{figure}

\begin{figure} 
\begin{center}
\includegraphics[width=8.5cm]{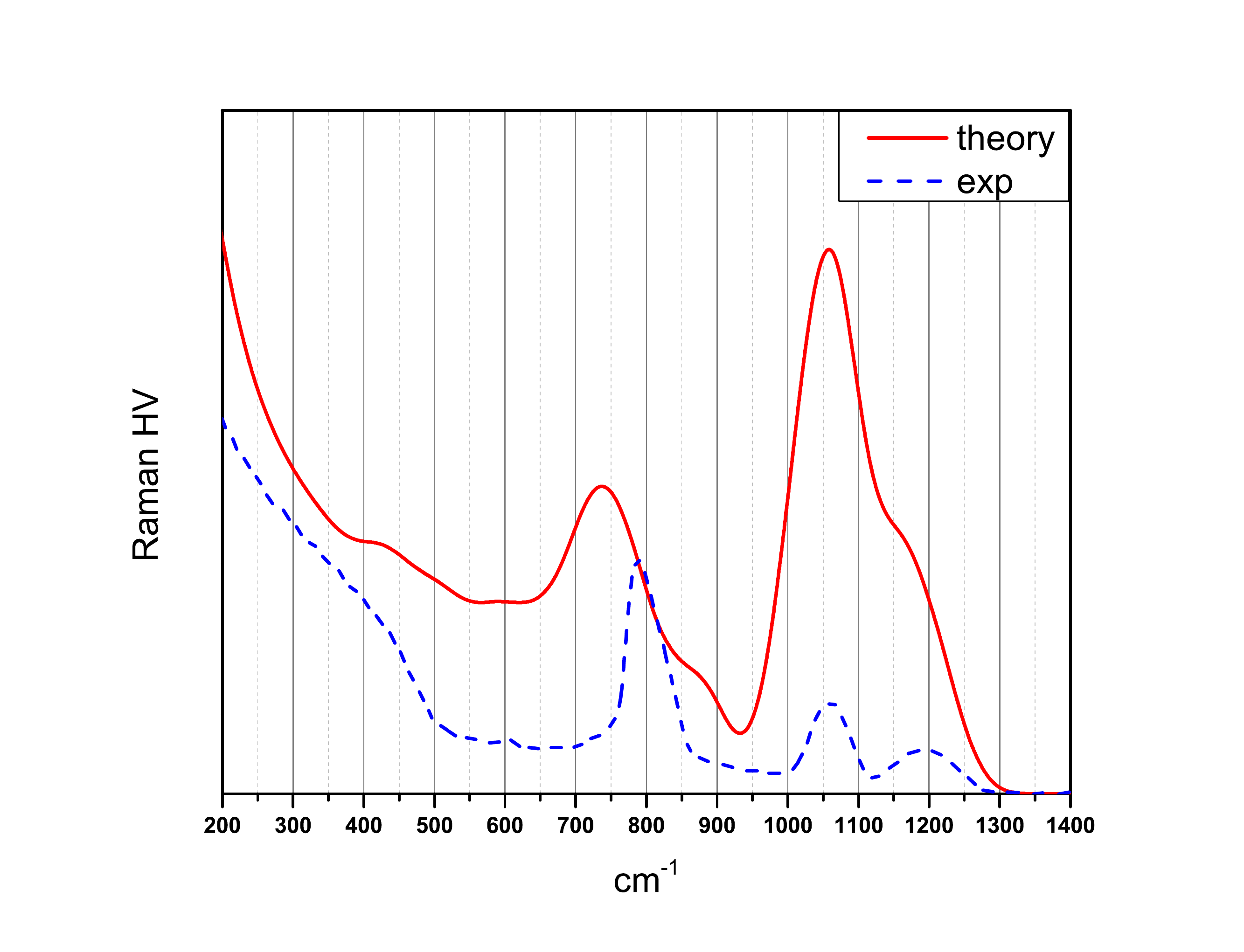}
\caption{ \label{figHVexp} Raman HV spectrum in uncompressed a-SiO$_{2}$. Red (solid) curve is theory [current work], black (dash) curve is experimental data from~\cite{sonneville2013}.}
\end{center}
\end{figure}

\begin{figure}  [!ht] 
\begin{center}
\includegraphics[width=8.5cm]{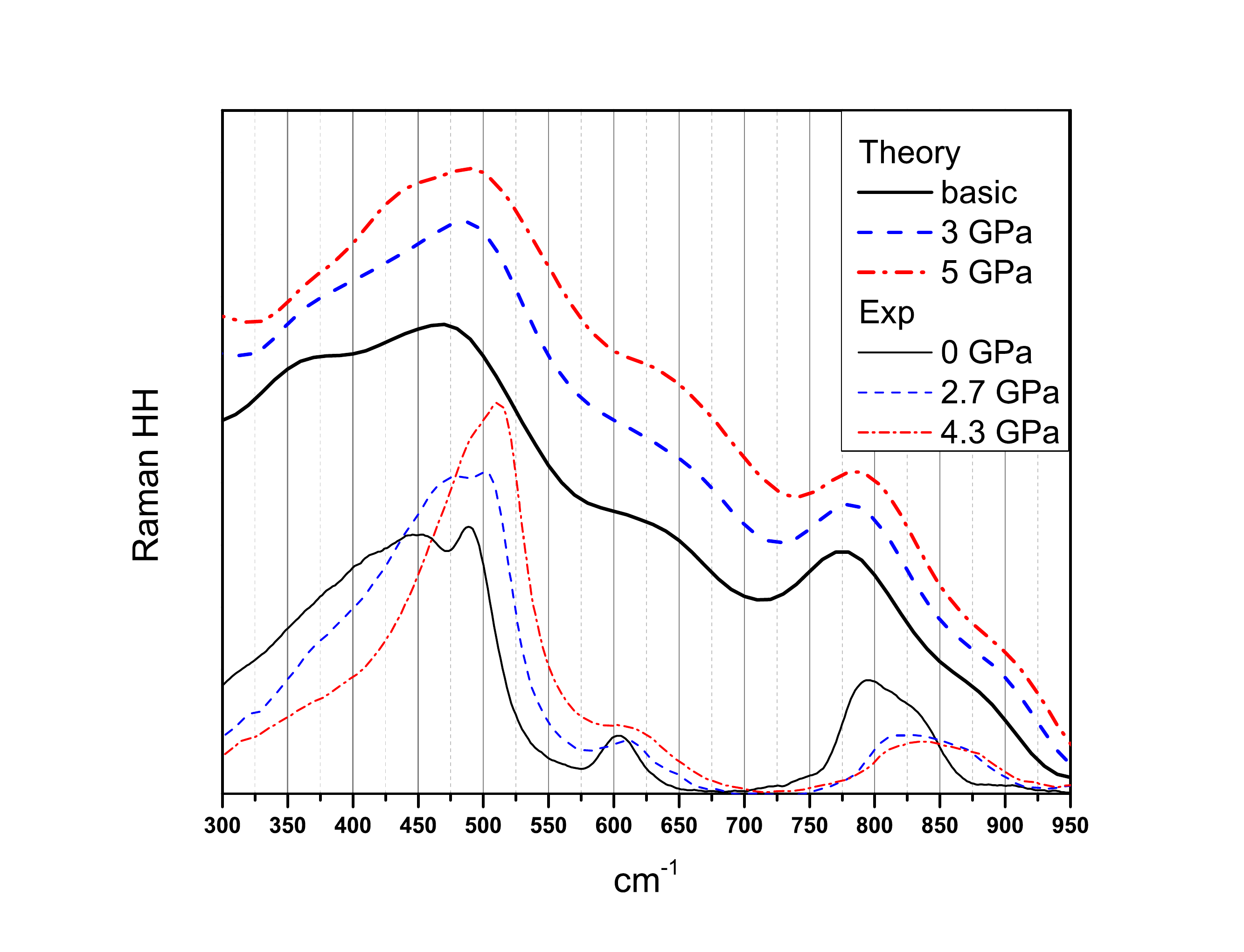}  
\caption{\label{figPexp} Raman HH spectrum in a-SiO$_{2}$ after an isotropic deformation has been applied, leading to different internal pressures P=0 GPa, P=3 GPa, and P=5 GPa (8232 particles), and comparison with experimental measurements from \cite{sonneville2013}. The curves have been shifted vertically by an arbitrary value for clarity. }
\end{center}
\end{figure}

We obtain the Raman HH and HV spectra by combining the classical eigenmodes obtained with the diagonalization of the (large scale) dynamical matrix and the BP model to describe the vibration dependence of the local polarizability. We computed the Raman spectra for the uncompressed samples (see Fig.~\ref{figHHexp} and~\ref{figHVexp}) and in the pressure imposed case (see Fig.~\ref{figPexp}) to validate our model. The results are  satisfying. Note that the spectra obtained in these figures have been obtained on a single sample without ensemble average. The computed spectrum was simply smoothened with gaussian functions of width 20 cm$^{-1}$ close to the experimental resolution. We see in Figures~\ref{figHHexp},~\ref{figHVexp}, and~\ref{figPexp} that the main characteristics of the experimental spectra are recovered within our semi-classical approximation.

In the parallel polarization case (HH spectrum), we can recognize well the main band ($\sim$ 400 - 550~cm$^{-1}$), and we can see, that the high-frequency experimental bands, in particular those located at $\sim$~800~cm$^{-1}$, $\sim$~1050~cm$^{-1}$, and $\sim$~1200~cm$^{-1}$ are well reproduced. The cross-polarized HV Raman spectrum gives similar results. These results are in good agreement with the experimental measurements~\cite{sonneville2013}, despite only the small shift of the 800~cm$^{-1}$ band to the left in the HV spectrum, with respect to the experimental data. In the pressure imposed case, the width and the position of the main band (400 - 550~cm$^{-1}$) has the same trend as in the experimental spectra (see Fig.~\ref{figPexp}): it shows a shift to higher frequencies and a progressive narrowing, that were the main results of~\cite{deschamps2009}. This successful comparison with experimental data encourages us to perform further analysis of our semi-classically computed Raman spectra for simple shear deformation of glasses, despite the lack of experimental data in this case.

\subsection{Shear sensitivity of Raman spectrum}

\begin{figure}   
\begin{center}
\includegraphics[width=8.5cm]{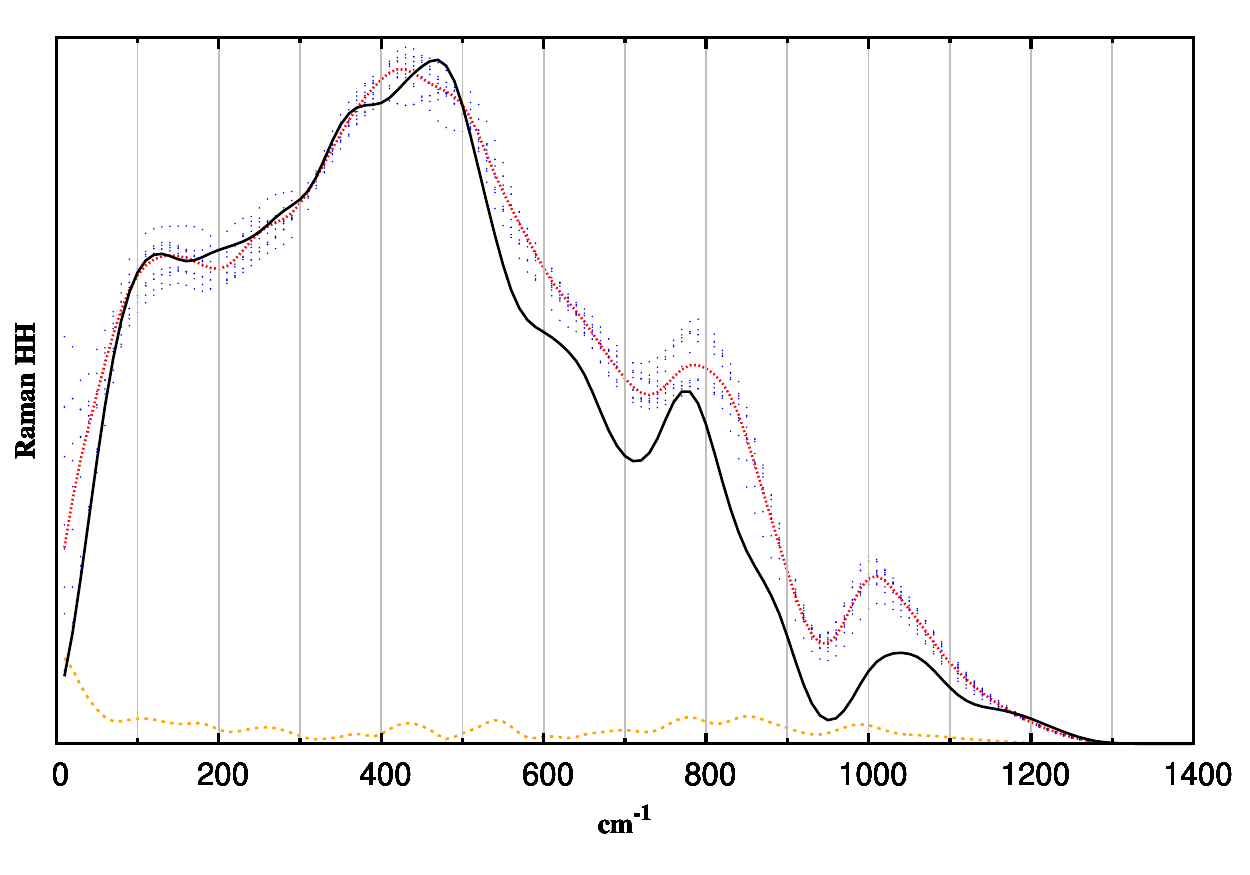}  
\caption{\label{figHHshear} Raman HH spectrum in a-SiO$_{2}$ upon shear. Black (solid) - basic sample, red (dash-dot) - mean spectrum, blue (doted area) - different configurations, orange (dot) - fluctuations.}
\end{center}
\end{figure}

\begin{figure} [!ht] 
\begin{center}
\includegraphics[width=8.5cm]{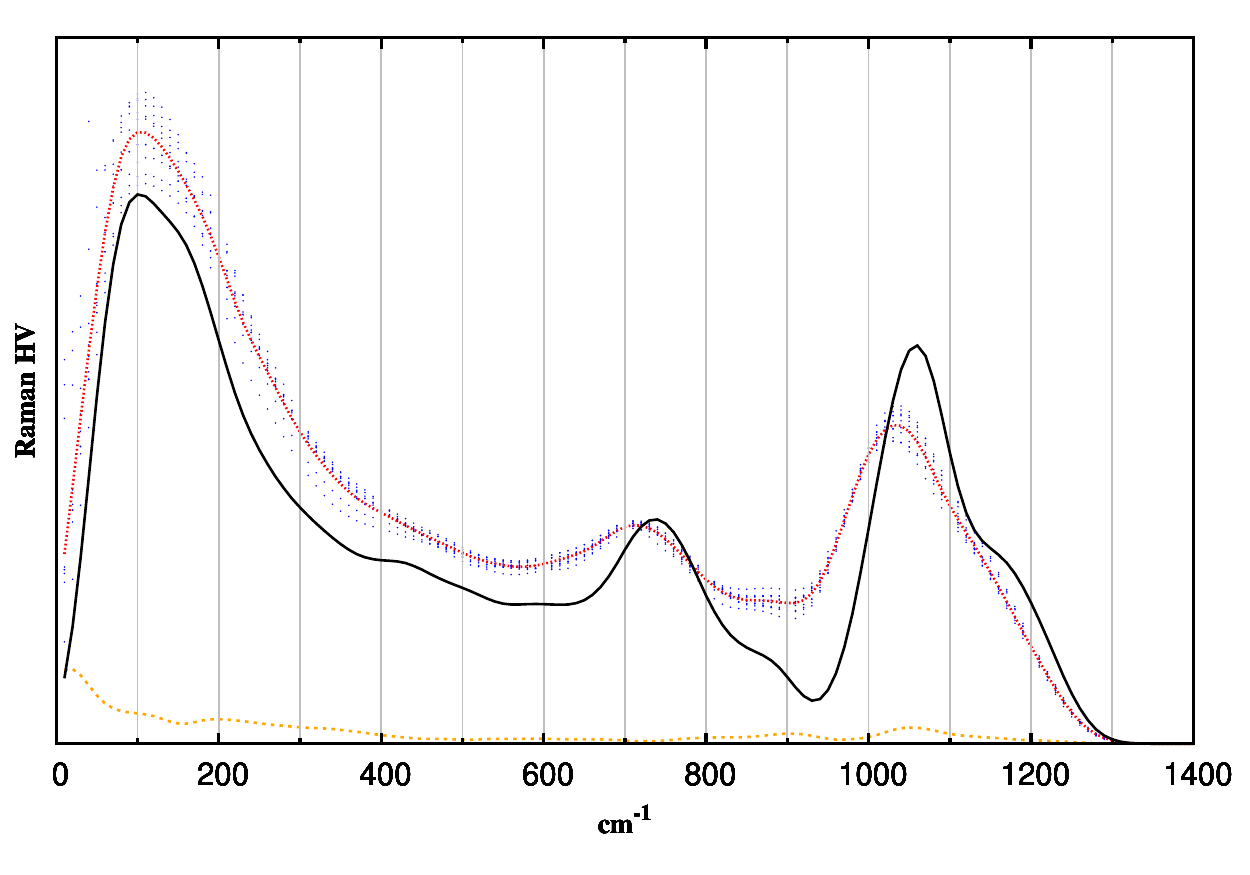}   
\caption{\label{figHVshear} Raman HV spectrum in a-SiO$_{2}$ upon shear. Black (solid) - basic sample, red (dash-dot) - mean spectrum, blue (doted area) - configurations, orange (dot) - fluctuations.}
\end{center}
\end{figure}

We mentioned above that the quasi-static deformation of our system is imposed through very fast energy relaxation after small imposed shear strain steps. The corresponding stress-strain behaviour is depicted in Fig.~\ref{figstress}. We are interested in the plastic regime, and more precisely in the plastic flow depicted in the {\it plastic plateau}. It seems reasonable to average over the configurations from the {\it plateau} of the stress-strain curve, since it is difficult to say what is the exact stage of our considered sample upon shear at constant stress in indentation experiment. The corresponding Raman HH and HV spectra, averaged over 10 configurations, are shown in Fig.~\ref{figHHshear} and~\ref{figHVshear}, respectively. Blue-dots is a coverage from contributions of different spectra which we average over, red (dash-dot) curve is the mean spectrum obtained on the sheared samples, and the black line is the spectrum of the initial unsheared sample. We see that the fluctuations (dotted curve) are not to large, that is an indisputable advantage of using larger samples in current work (8232 particles) with respect to most past works, where smaller ($\sim$ 1000 particles) samples could be analyzed hardly due to finite size effect~\cite{smirnov2008}. In our semi-classical simulations of Raman spectra performed on sufficiently large systems, it appears easy to separate the effect of shear from the configurational noise.

Both HH and HV spectra show clearly the activation of high-frequency modes upon shear, whereas the low-frequency domain (below 500 cm$^{-1}$) remains without valuable changes. Both spectra show an increase of the Raman activity in the range 800-1000 cm$^{-1}$ as does the density of vibrational states (Fig.~\ref{fig-DOS-O-Si}), and a shift of the 1050 cm $^{-1}$ band to lower frequencies. An analysis of the high-frequency mode contributions is not obvious \cite{luigi2007}, therefore we must be careful to treat this range. Our simulations show that the plastic shear flow induces irreversible structural changes in the model silica glass studied here, with a transformation of large rings into lower order units. The concomitant increase of the D2-band could support the usual interpretation in terms of vibrations of 3-fold rings~\cite{galeener85}. 

At larger frequencies, the range $\sim$ 800 - 1000~cm$^{-1}$ has an obvious growth for HV \& HH spectra. Particularly, the HV spectrum shows clearly this growth, whereas the bands $\sim$ 1050~cm$^{-1}$ and $\sim$ 1200~cm$^{-1}$ become less expressed with shear but their contribution to the spectra remains. Note that the Raman signal is not sufficiently sensitive to discriminate between the configurations of silica in the plastic plateau that are close to a plastic instability, and those that are deeply stable, unlike in~\cite{tanguy2010}. This is probably due to the fact that Raman signals are more sensitive to the irreversible structural changes that appeared in the beginning of the plastic plateau, than to the sparse mechanical instabilities responsible for the small scale stress fluctuations in the plateau. We will now relate this behaviour to a detailed study of the shear sensitivity in the density of vibrational states, and then to a detailed study of the vibration modes at different eigenfrequencies. 

\subsection{Density of Vibrational States}

\begin{figure}
\begin{center}
\includegraphics[width=8.5cm]{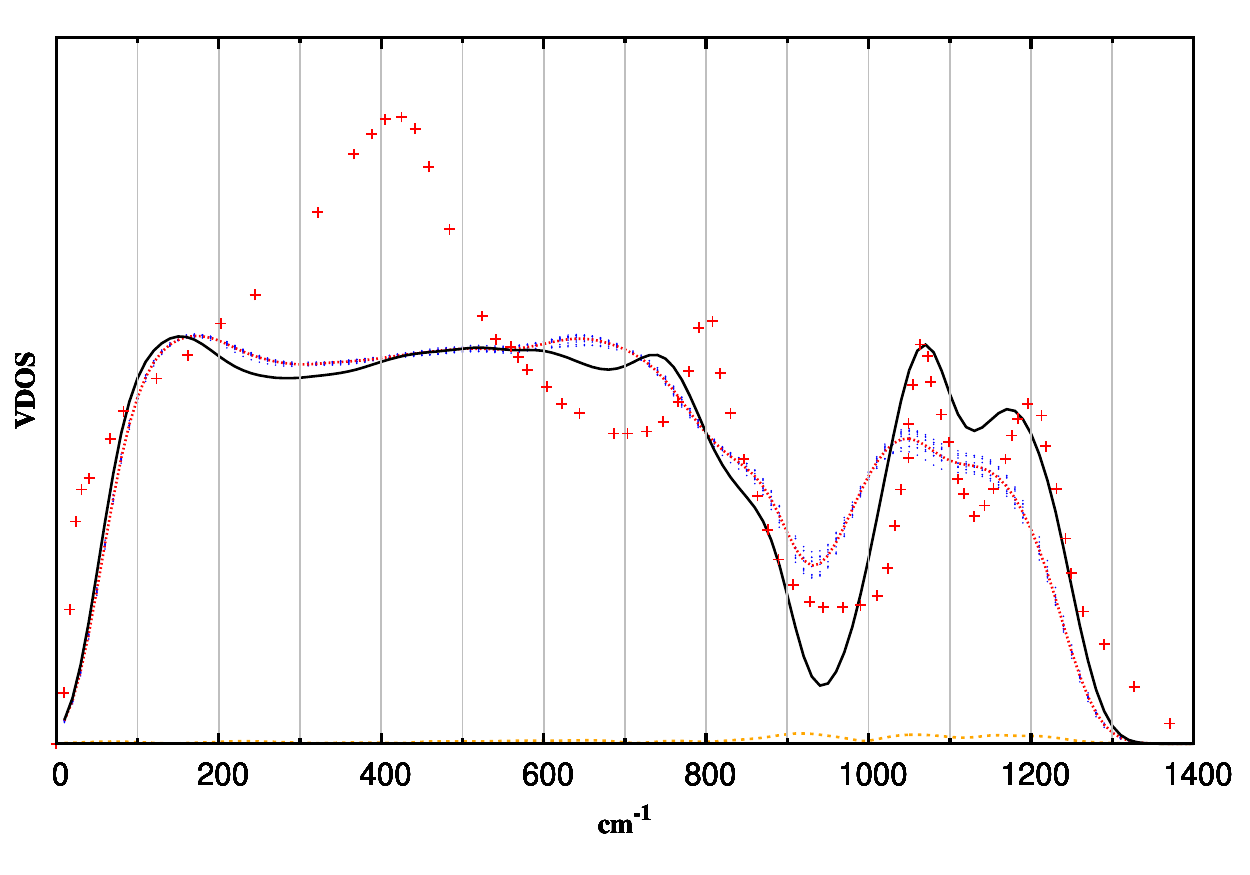}
\caption{ \label{fig-DOS-shear} VDOS spectrum in a-SiO$_{2}$ upon shear. Black (solid) - basic sample, red (dash-dot) - mean spectrum, blue (doted area) - configurations, orange (dot) - fluctuation, cross - experimental results from~\cite{price1987}.}
\end{center}
\end{figure}

\begin{figure} 
\begin{center}
\includegraphics[scale=.34]{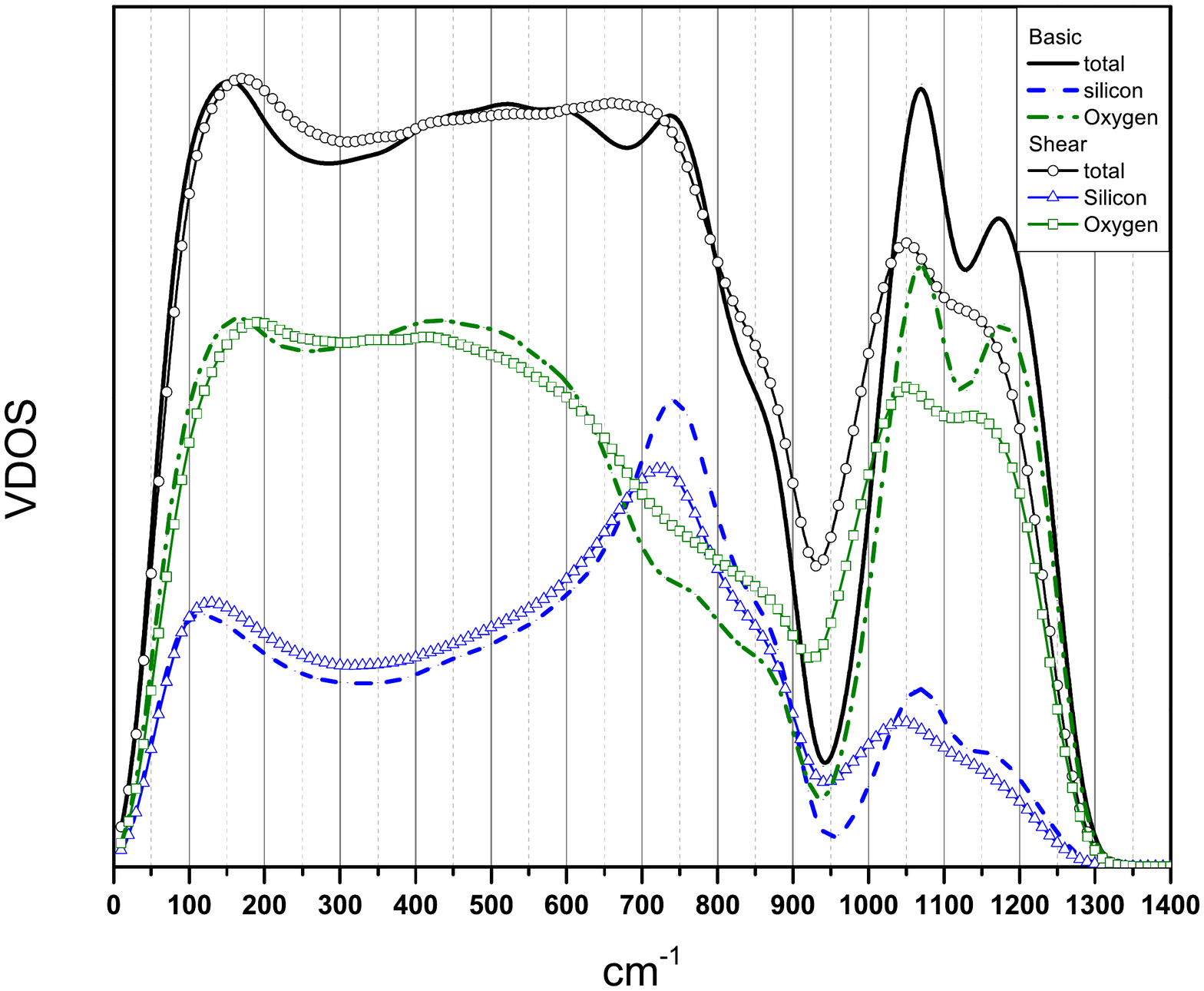}
\caption{\label{fig-DOS-O-Si} Basic vs Shear of VDOS spectra in a-SiO$_{2}$ with the different contributions of Si and O atoms to the vibrations.}
\end{center}
\end{figure}

\begin{figure} 
\begin{center}
\includegraphics[scale=.34]{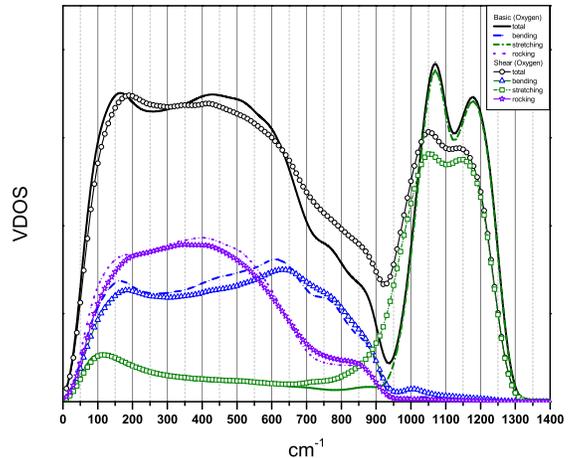}
\caption{\label{fig-DOS-B-R-S} Basic vs Shear of VDOS spectra in a-SiO$_{2}$ with the different contributions of bending, stretching, and rocking types of vibrations. The total corresponds to the oxygen contribution shown in Fig.~\ref{fig-DOS-O-Si}}
\end{center}
\end{figure}

\begin{figure} 
\begin{center}
\includegraphics[scale=.34]{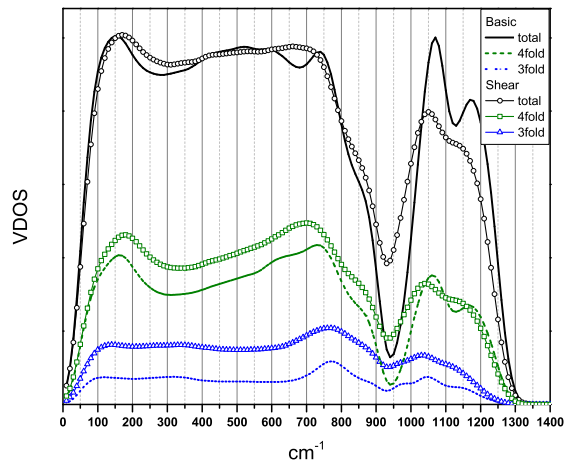}
\caption{\label{fig-DOS-3f-4f} Basic vs Shear of VDOS spectra in a-SiO$_{2}$ with the different contributions of 3-fold and 4-fold rings.}
\end{center}
\end{figure}

\begin{figure} 
\begin{center}
\includegraphics[scale=.34]{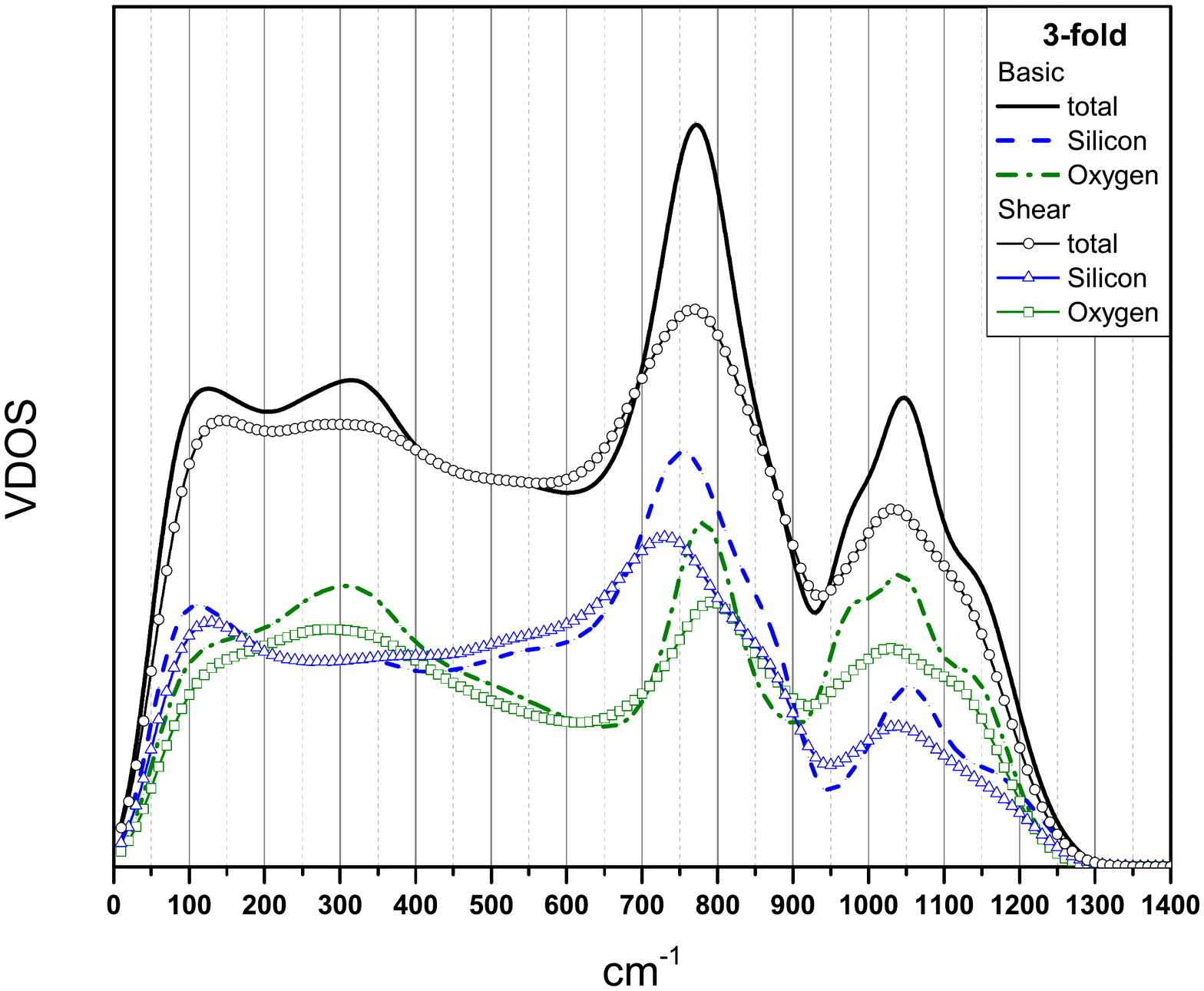}
\caption{\label{fig-DOS-3f-O-Si} Basic vs Shear of VDOS spectra in a-SiO$_{2}$ with the different contributions of Si and O atoms to the vibrations supported by 3-fold rings.}
\end{center}
\end{figure}

\begin{figure} 
\begin{center}
\includegraphics[scale=.35]{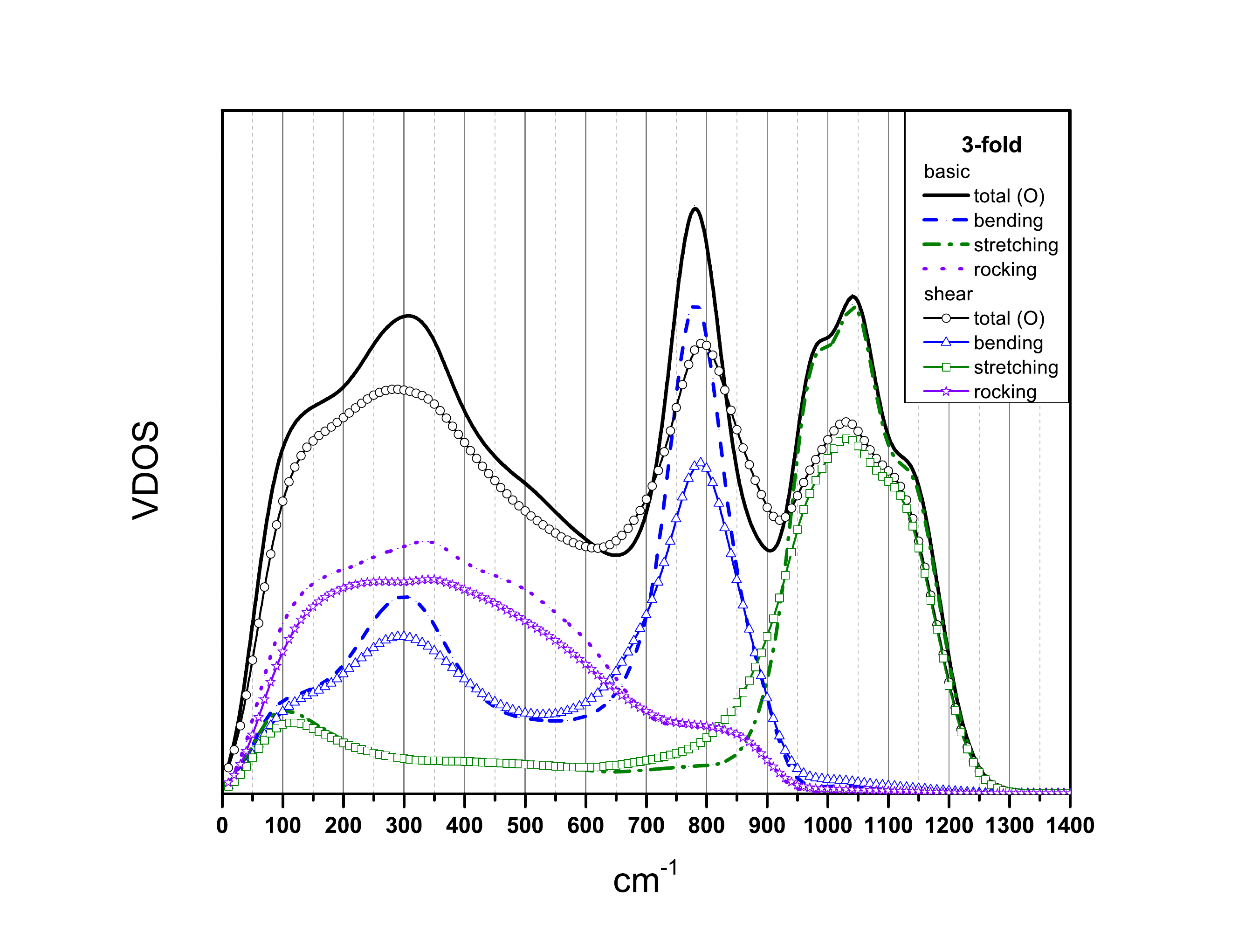}
\caption{ \label{fig-DOS-3f-B-S-R} Basic vs Shear of VDOS spectra in a-SiO$_{2}$ with the different contributions of bending, stretching, and rocking types of oxygen atom vibrations supported by 3-fold rings.}
\end{center}
\end{figure}

\begin{figure} 
\begin{center}
\includegraphics[scale=.34]{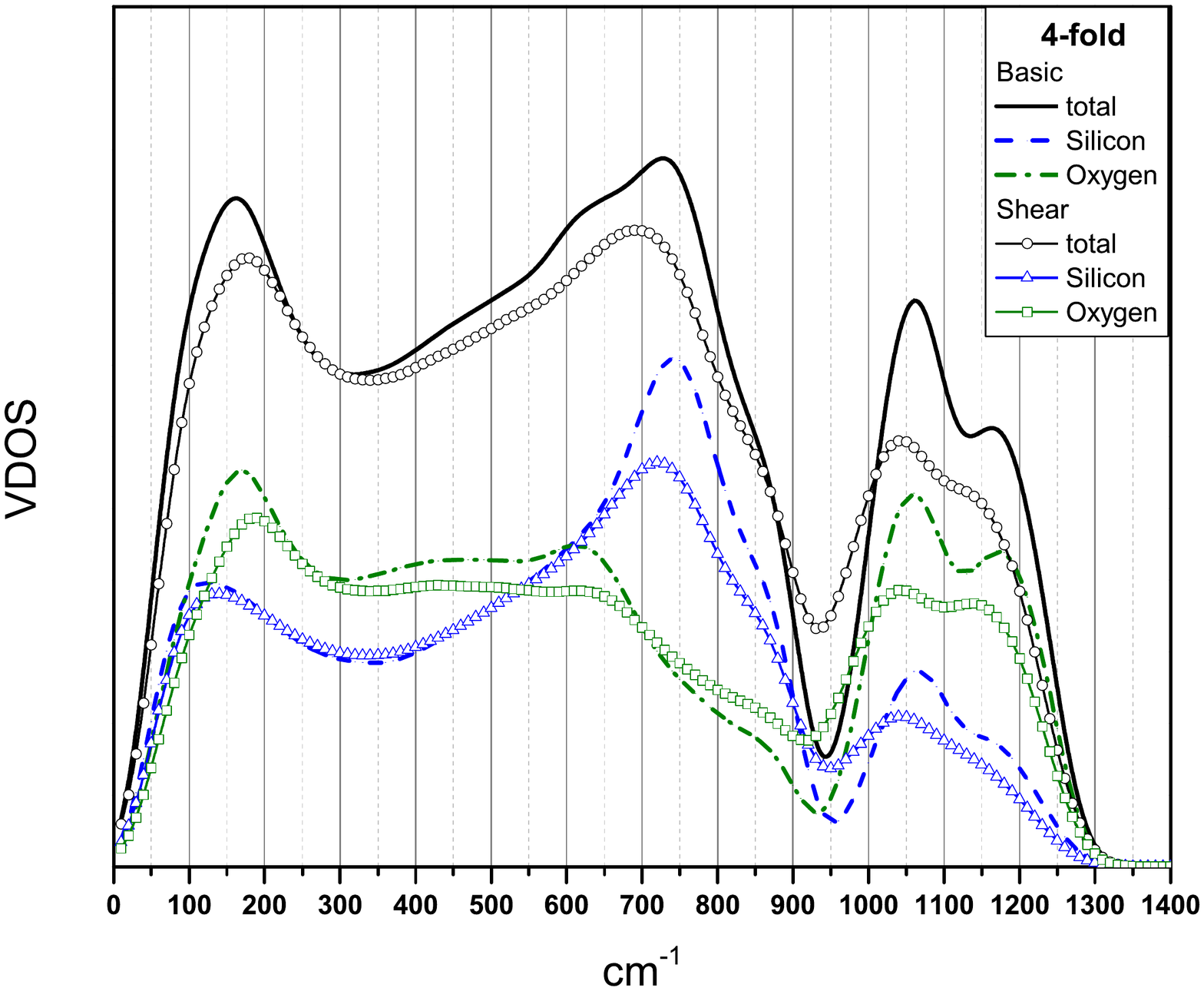}
\caption{\label{fig-DOS-4f-O-Si} Basic vs Shear of VDOS spectra in a-SiO$_{2}$ with the different contributions of Si and O atoms to the vibrations supported by 4-fold rings.}
\end{center}
\end{figure}

\begin{figure} 
\begin{center}
\includegraphics[scale=.34]{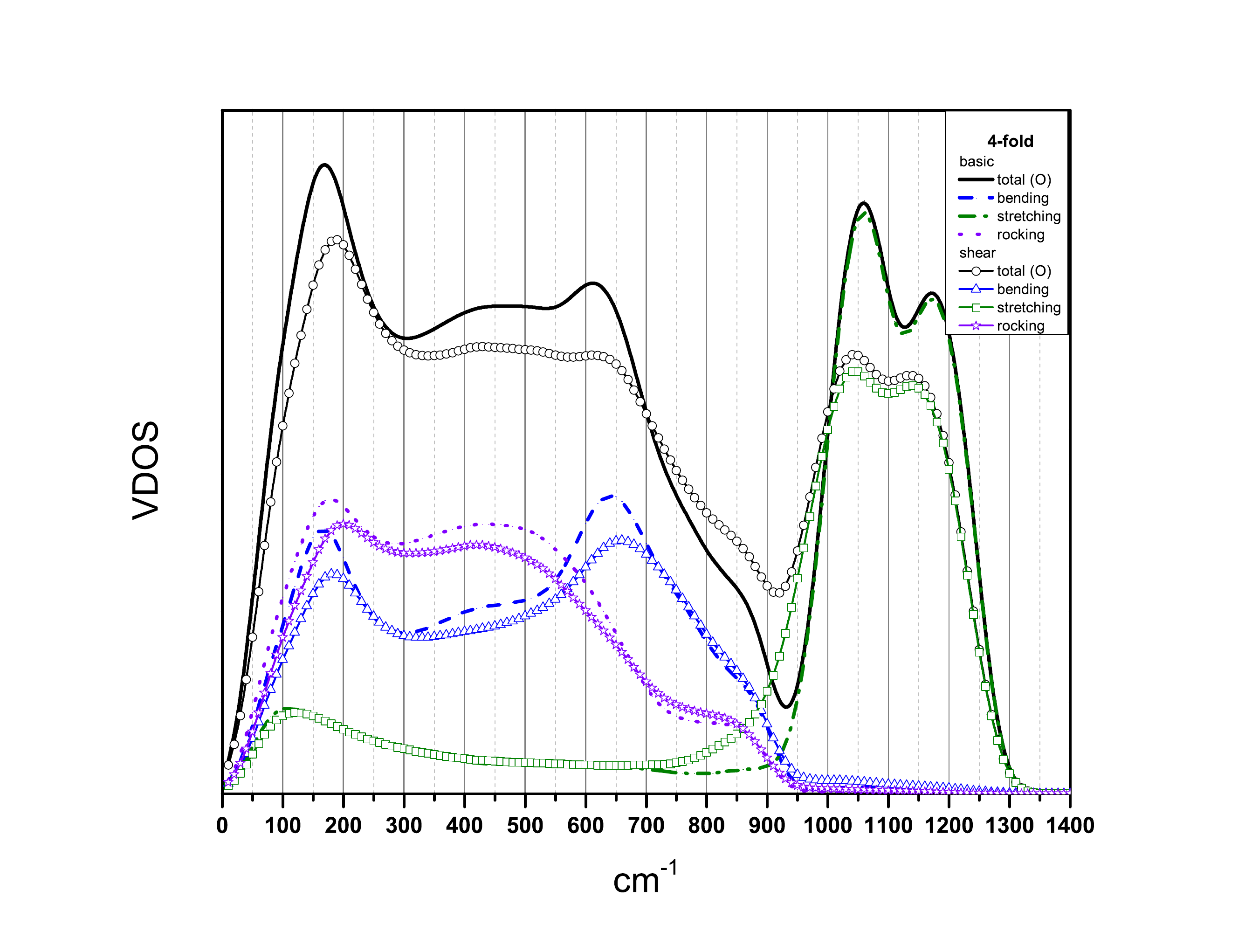}
\caption{\label{fig-DOS-4f-B-S-R} Basic vs Shear of VDOS spectra in a-SiO$_{2}$ with the different contributions of bending, stretching, and rocking types of oxygen atom vibrations supported by 4-fold rings.}
\end{center}
\end{figure}

\begin{figure}  
\begin{center}
\includegraphics[width=8.5cm]{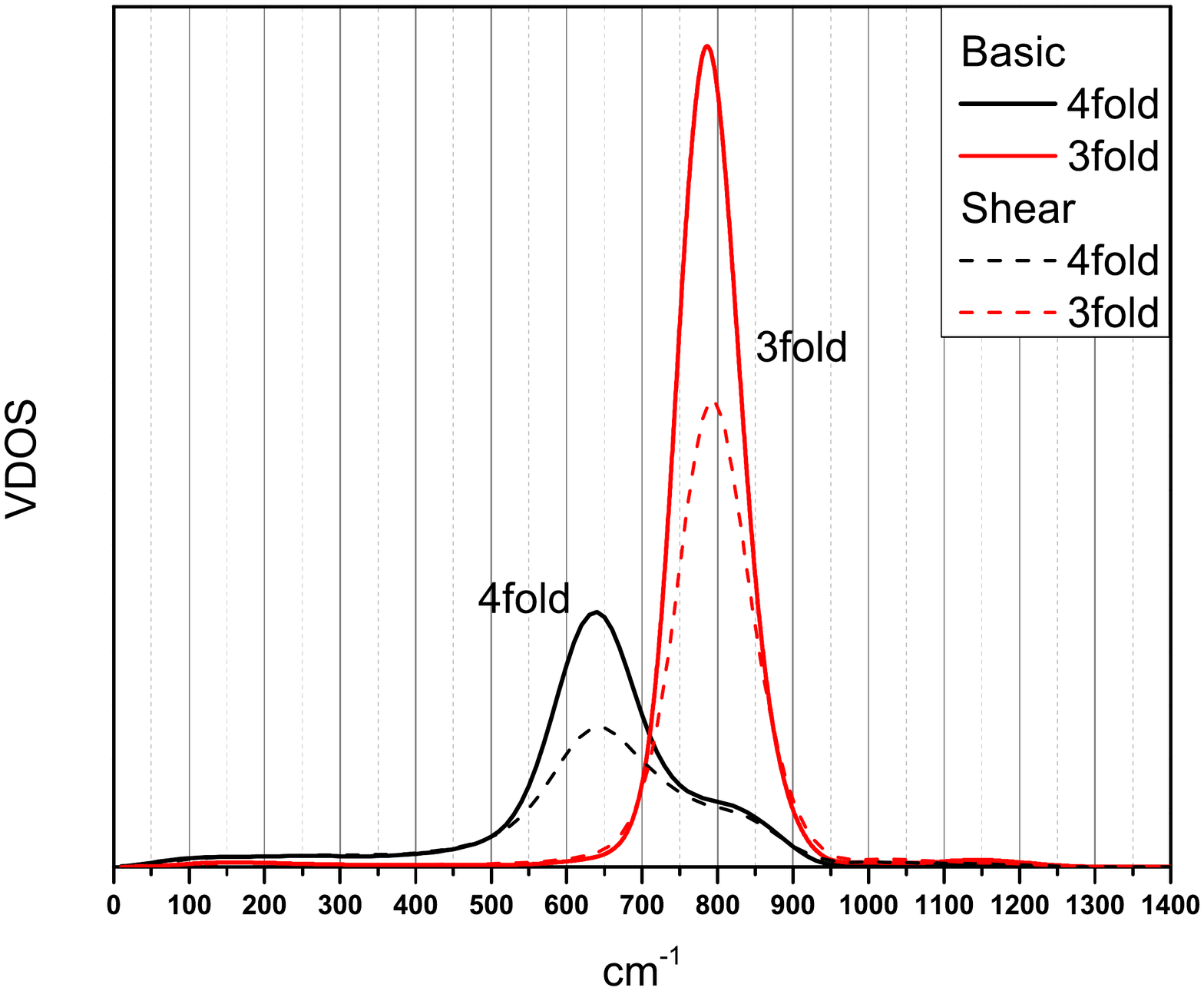}
\caption{\label{fig-BREATHING} Basic vs Shear contribution to the VDOS of in-phase breathing modes supported by oxygen atoms in 3-fold and 4-fold rings, divided by the corresponding number of rings.}
\end{center}
\end{figure}

Following the same logic as before for Raman, the Vibrational Density of States (VDOS) can be evaluated with the help of Eq.~\eqref{ex22}. We can see in Fig.~\ref{fig-DOS-shear} the changes appearing in the VDOS after the system entered in the plastic flow regime. These changes are mainly located in the high-frequency domain ($> 1000$~cm$^{-1}$) and in the intermediate frequency range (700~cm$^{-1}$-1000~cm$^{-1}$), that correlate with the Raman spectra. Due to the activation of the VDOS in the vicinity of $\sim$ 950 cm$^{-1}$ upon shear, the populations of Raman HH and Raman HV spectra have the same trend. The changes in the VDOS are related to a redistribution of the vibrational activity: on one hand we observe the decrease of population in the $\sim$ 500 and $\sim$ 1000-1200 cm$^{-1}$ bands, on the other hand an activation of the mute area between these bands occurs, as well as an enhancement in the range of D2-band ($\approx 600$~cm$^{-1}$). 

The analysis of the partial contributions from Si-/O-atoms to VDOS (see Fig.~\ref{fig-DOS-O-Si}) is very instructive. In agreement with previous results obtained in Refs.~\cite{jin1993,taraskin1997} on different simulation of silica samples, it shows that the high frequency ($>$ 950~cm$^{-1}$) and the low frequency domains ($<$ 650~cm$^{-1}$) for the samples at rest, are dominated by the vibrations of oxygen atoms. A peak in the contribution of Si displacements to the VDOS is located at $\sim$730~cm$^{-1}$. Upon shear, the decay of the high frequency VDOS is univocally related to the decay in the oxygen motion, with a poorer definition and a small shift of the two peaks located at 1050~cm${-1}$ and 1200~cm$^{-1}$. Similarly, the increase of the activity in the intermediate range 700-1000~cm$^{-1}$ is due mainly to the increase of the oxygen displacements in that frequency range.

Finally, the decomposition of the modes into bending/stretching/rocking modes (see Fig.~\ref{fig-DOS-B-R-S}) following the Eqs.~\eqref{eq-R}~-~\eqref{eq-B} shows in agreement with~\cite{taraskin1997,alf1998,giustino2006,luigi2009} that rocking and bending modes dominate at low frequencies, while the high frequency range is dominated by stretching modes. Upon shear, we show in this figure, that the increase of oxygen activity in the intermediate domain is uniquely due to the increasing number of stretching modes in that frequency range

In order to compare with the usual interpretation of high-frequency vibrations, we will now study in more details the contribution of the different rings to the vibrations, and its dependence with the plastic shear deformation. Fig.~\ref{fig-DOS-3f-4f} shows the contribution of 3-fold and 4-fold rings to the VDOS. The general increase of 3-fold and 4-fold contribution upon shear, is due to the increasing number of such rings upon plastic shear (Fig.~\ref{figrings-shear}). In the high frequency domain, the contribution of 4-fold rings is stable, but the increase of the contribution of 3-fold ring is opposite to the decay of the total VDOS in this frequency range. This proves that the decay of the oxygen displacements responsible for the decay of the total activity at high frequency is located outside the small rings. 

Let's first focus on the contribution of 3-fold rings to the activity. Fig.~\ref{fig-DOS-3f-O-Si} shows the partial contribution of 3-fold rings to VDOS, normalized by the number of rings. The main contribution of 3-fold rings appears at $\sim$770~cm$^{-1}$ (thus at a frequency higher than the frequency of the D2 band). This peak is mainly due to the contribution of bending modes (Fig.~\ref{fig-DOS-3f-B-S-R}) and it is smoothened upon shear, for oxygen as well as for silicon atoms. Oxygen atoms show however an increasing activity in the intermediate frequency range (700 - 1000 cm$^{-1}$). This increase of the activity in the intermediate frequency range is related to an increase of stretching modes, as already seen for the whole system.

Let's consider now 4-fold rings. The main contribution of 4-fold rings is located at $\sim$700~cm$^{-1}$ and is due tot the enhanced vibration of silicon atoms at this frequency (Fig.~\ref{fig-DOS-4f-O-Si}). This peak is smoothened upon shear, but there is a concomitant increase of the oxygen activity in the intermediate frequency range (700 - 1000 cm$^{-1}$) upon stretching modes, that contributes to the general increase of VDOS in the intermediate frequency range. Note interestingly, that the main contributions of oxygen are located at 150~cm$^{-1}$ and at 630~cm$^{-1}$, and do not correspond to D1 band. 

\begin{figure} 
\begin{center}
\includegraphics[width=8.5cm]{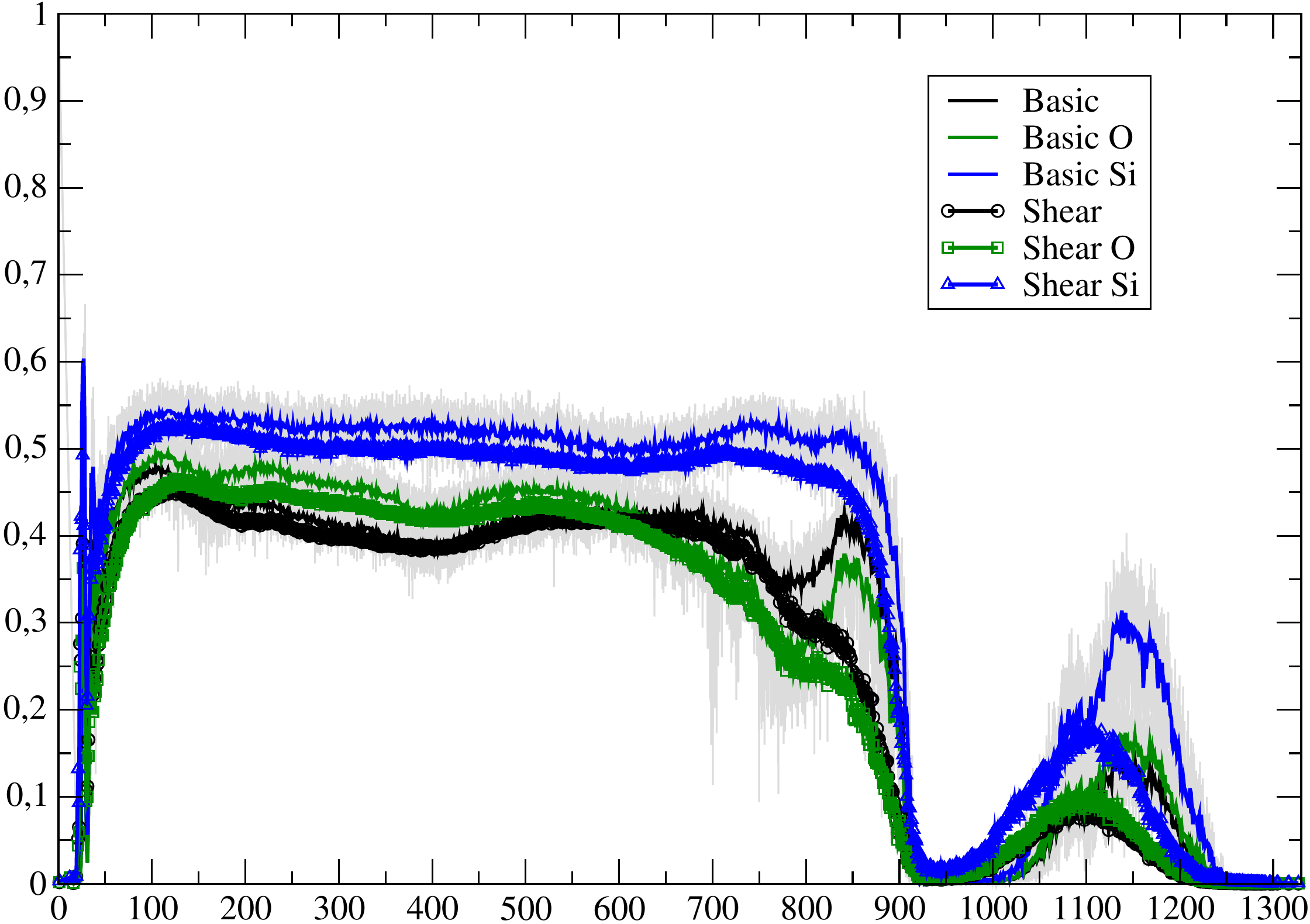}
\caption{\label{fig-TP-shear} Basic vs Shear Participation Ratio in a-SiO$_{2}$, as a function of the frequency, and partial Participation Ratios restricted to O and Si atoms.}
\end{center}
\end{figure}

\begin{figure}
\begin{center}
\includegraphics[width=4cm]{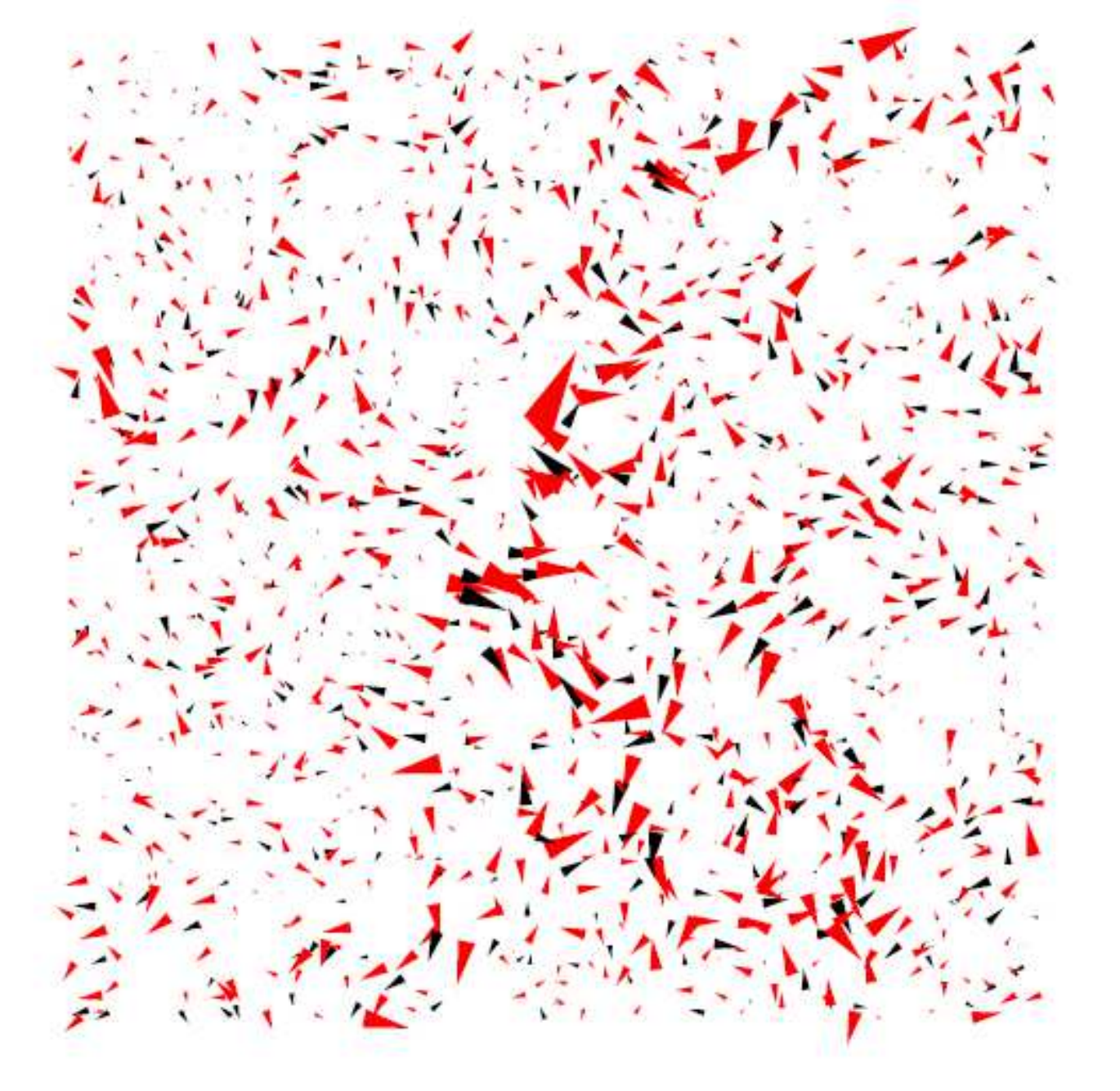}
\includegraphics[width=4cm]{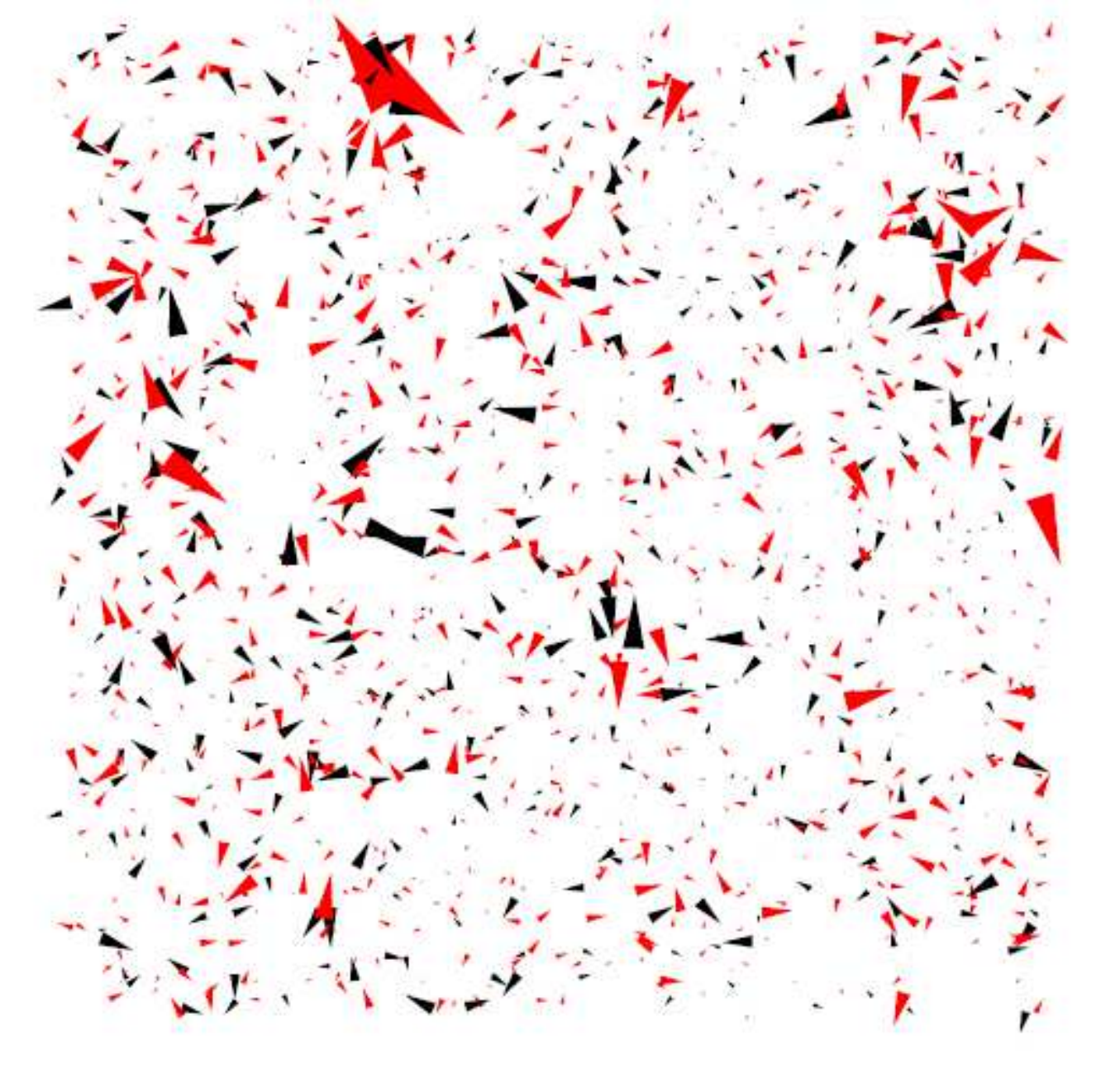}
\caption{\label{fig-mode-plateau} (a) Vibration mode at the frequency 97~cm$^{-1}$ for the basic configuration and (b) at the frequency 849~cm$^{-1}$. The arrows are proportional to the atomic displacements. Red arrows are for oxygen atoms, and Black arrows for silicon atoms.}
\end{center}
\end{figure}

In order finally to test the interpretation given by Galeener~\cite{galeener85, giustino2006} for D1 and D2 band in our samples, we have isolated the contribution of in-phase bending vibrations of oxygen atoms in 3-fold and 4-fold rings, that correspond to the breathing vibration modes of the rings. The contribution of these modes to the VDOS is shown in Fig.~\ref{fig-BREATHING}. It is remarkable that we see two well defined peaks. It should be noted that the in-phase breathing modes of Oxygen atoms in 4- and 3-fold rings have been attributed respectively to the defect lines D1 and D2 in the HH Raman spectrum at 495 and 605 cm$^{-1}$. It is not a surprise that we find the two lines at higher frequencies as similar overestimations~\cite{to2007} are found when the BKS potential is used for evaluating vibrational modes of silica polymorphs for which experimental data are known. However, we show here that we can predict the behaviour of D1 and D2 lines in the Raman HH spectrum upon an external perturbation such as shear. Indeed, D1 and D2 lines are expected to increase upon shear as can be seen in Fig.~\ref{figHHshear} keeping in mind the we assign the theoretical D1 line to the feature at $\sim$630~cm$^{-1}$ and the D2 line to that at  $\sim$770~cm$^{-1}$ . 

\subsection{Analysis of the Vibration Modes}

The vibration modes are the resonant eigenmodes obtained by the exact diagonalization of the Dynamical Matrix, as explained in Part \ref{part:RamanModel}.

A parameter that is frequently used to characterize the vibration modes in amorphous systems is the Participation Ratio (PR). It is defined as 
\begin{equation}
PR=\frac{1}{N}\frac{\left(\sum_{i=1}^{N} {{\left | {\bf u}_i \right |}^2} \right)^2}{\sum_{i=1}^{N}{ {\left | {\bf u}_i \right |}^4}}.
\label{eq:PR}
\end{equation}

\begin{figure}  [!t]
\begin{center}
\includegraphics[width=8.5cm]{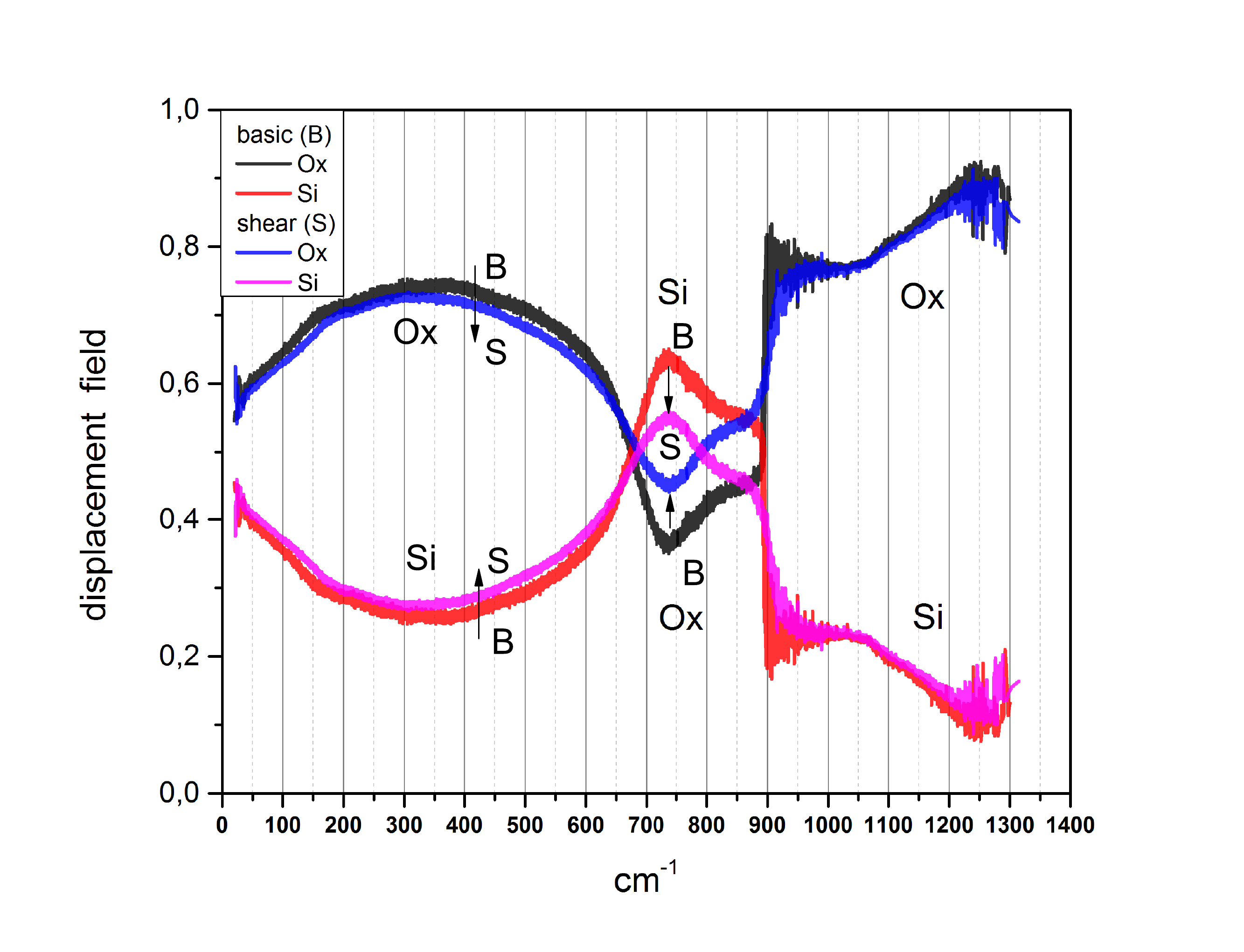}
\caption{\label{fig-DISP-O-Si} Contribution of Si vs O atoms to the displacement field in the eigenvectors, as a function of the mode's eigenfrequency. Comparison upon steady shear.}
\end{center}
\end{figure}

\begin{figure} 
\begin{center}
\includegraphics[width=8.5cm]{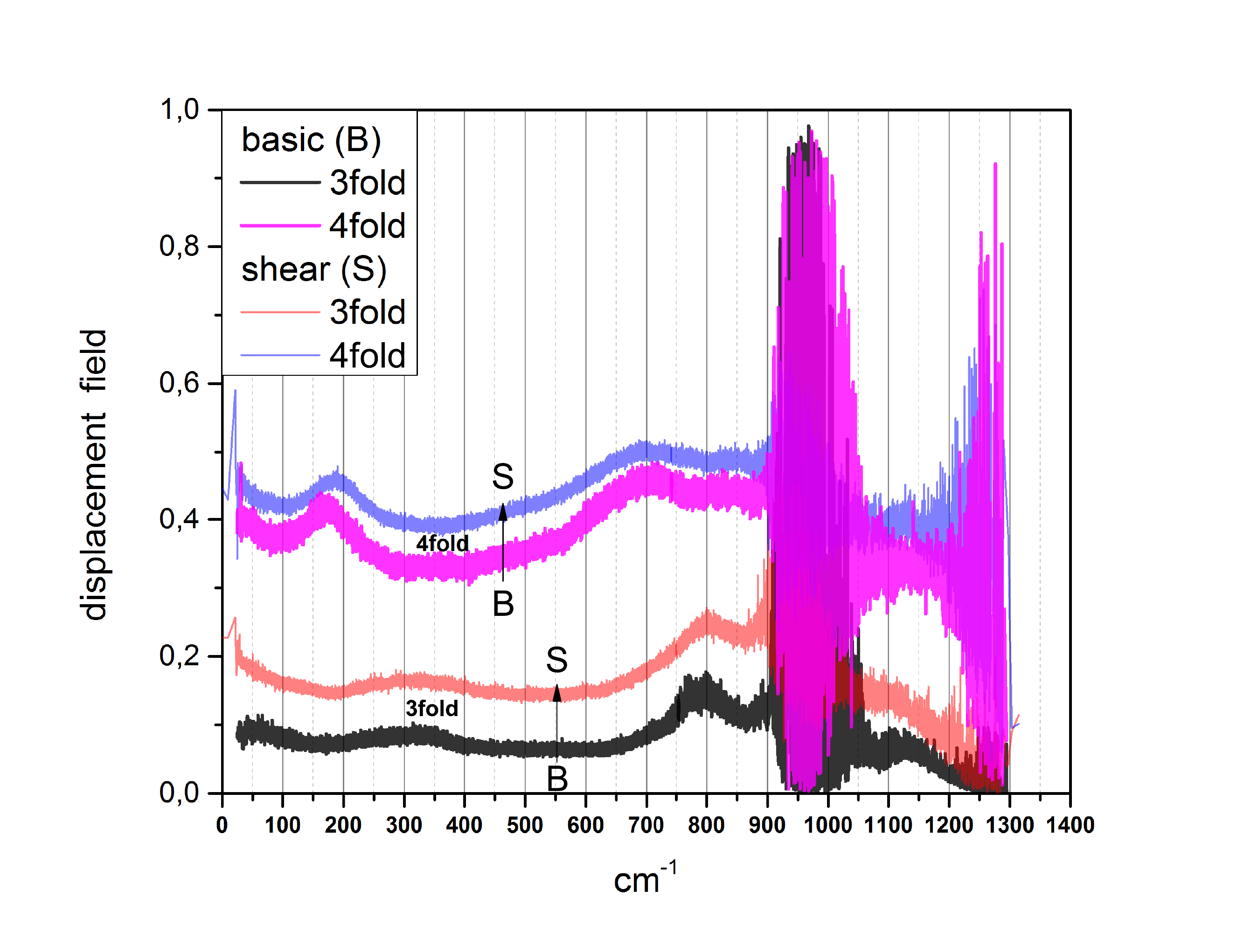}
\caption{ \label{fig-DISP-3-4} Contribution of 3- \& 4-fold rings to the displacement field in the eigenvectors as a function of the mode's eigenfrequency, and effect of shear.}
\end{center}
\end{figure}

It quantifies the percentage of particles moving with the same amplitude ($PR=1/N$ if one single particle is moving, $PR=1$ if the $N$ particles are moving together), and is often used to characterize the degree of localization of the modes. The Participation Ratios of the vibration modes obtained from the eigenvectors of the dynamical matrix are depicted in Fig.~\ref{fig-TP-shear}. It shows different frequency regimes: the first 3 translational modes in each of the 3 directions have a Participation Ratio, $PR=1$. At very low frequencies ($<$ 50~cm$^{-1}$), the Participation Ratio of the plane waves decays with the frequency (smaller wavelength). In the same low frequency regime ($<$ 50~cm$^{-1}$), we can identify soft modes that are characterized by a very low Participation Ratio. These modes correspond to localized vibrations with unusual low frequencies, that are precursors of local plastic instabilities~\cite{tanguy2010}. Surprisingly enough, in silica glasses, these modes are present whatever the distance to a plastic instability, unlike in other glassy systems where they appear only at the onset of plasticity~\cite{tanguy2010}. We will not study them in more details here, because they are not in the frequency range where the main changes to the Raman spectra appear upon plastic deformation. At higher frequencies, close to 100~cm$^{-1}$, the Participation Ratio reaches a local maximum, followed by a plateau. The frequency where this local maximum appears (with $PR\approx 0.45$) is close to the Boson's peak frequency obtained by dividing the VDOS by the Debye prediction~\cite{Boris2012}. At this frequency, the modes display large scale rotational motions structures (see Fig.~\ref{fig-mode-plateau}-a). In the plateau, the rotational motion is more noisy. At higher frequencies, the Participation Ratio shows a small decay and a local maximum ($\omega\sim$850~cm$^{-1}$) followed by a strong decay at a frequency close to $900$~cm$^{-1}$. The global Participation Ratio discussed here, can be compared to the partial Participation Ratios, where the sum in Eq.~\eqref{eq:PR} runs only over one type of atoms (Si or O). It is seen in Fig.~\ref{fig-TP-shear}, that the Participation Ratio restricted to oxygen atoms is very close to the global one. The Participation Ratio of silicon atoms does not depend to much on the frequency and is quite high, indicating homogeneous motion of the silicon skeleton. The variation of the Participation Ratio near 850~cm$^{-1}$ upon shear is directly related to a change in the spatial distribution of oxygen displacements, that start to localize smoothly at 700~cm$^{-1}$ while the silicon matrix still vibrates collectively ($PR_{Si}\sim 0.55$ but $PR_{O}\sim 0.25$ at 800~cm$^{-1}$). Oxygen motion appears thus very sensitive to the frequency in the intermediate frequency range 700-900~cm$^{-1}$. After a strong decay of the Participation Ratio corresponding to localization processes~\cite{Tanguy2014}, the PR increases again between $1000$ and $1200$~cm$^{-1}$. As seen before, these high frequencies correspond to local stretching modes\cite{alf1997}. The main effect of plastic shear is to smoothen strongly the bump in the participation ratio close to $850$~cm$^{-1}$. This effect is related to a partial localization of the modes~supported by oxygen atoms in this frequency range upon plastic shear, and will be discussed later in more details. 

\begin{figure} [t]
\begin{center}
\includegraphics[width=8.5cm]{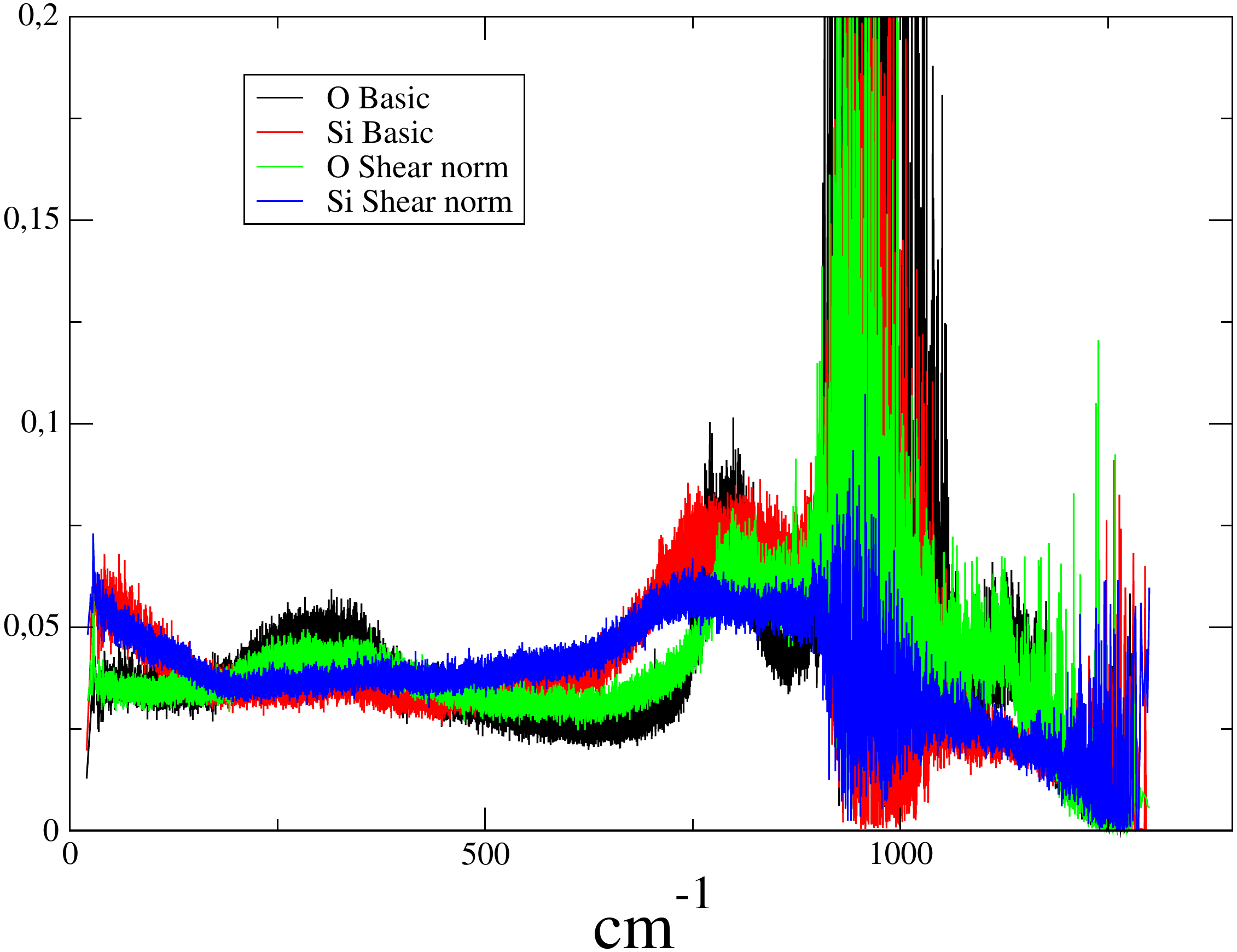}
\caption{\label{fig-DISP-3-O-Si} Separate contribution of Si \& O atoms in the 3-fold rings, to the displacement field in the eigenvectors as a function of the mode's eigenfrequency, and effect of shear. The curves corresponding to the sheared state have been normalized by the ratio between the number of 3-fold rings in the sheared state, and the number of 3-fold rings in the initial state.}
\end{center}
\end{figure}

\begin{figure} [t]
\begin{center}
\includegraphics[width=8.5cm]{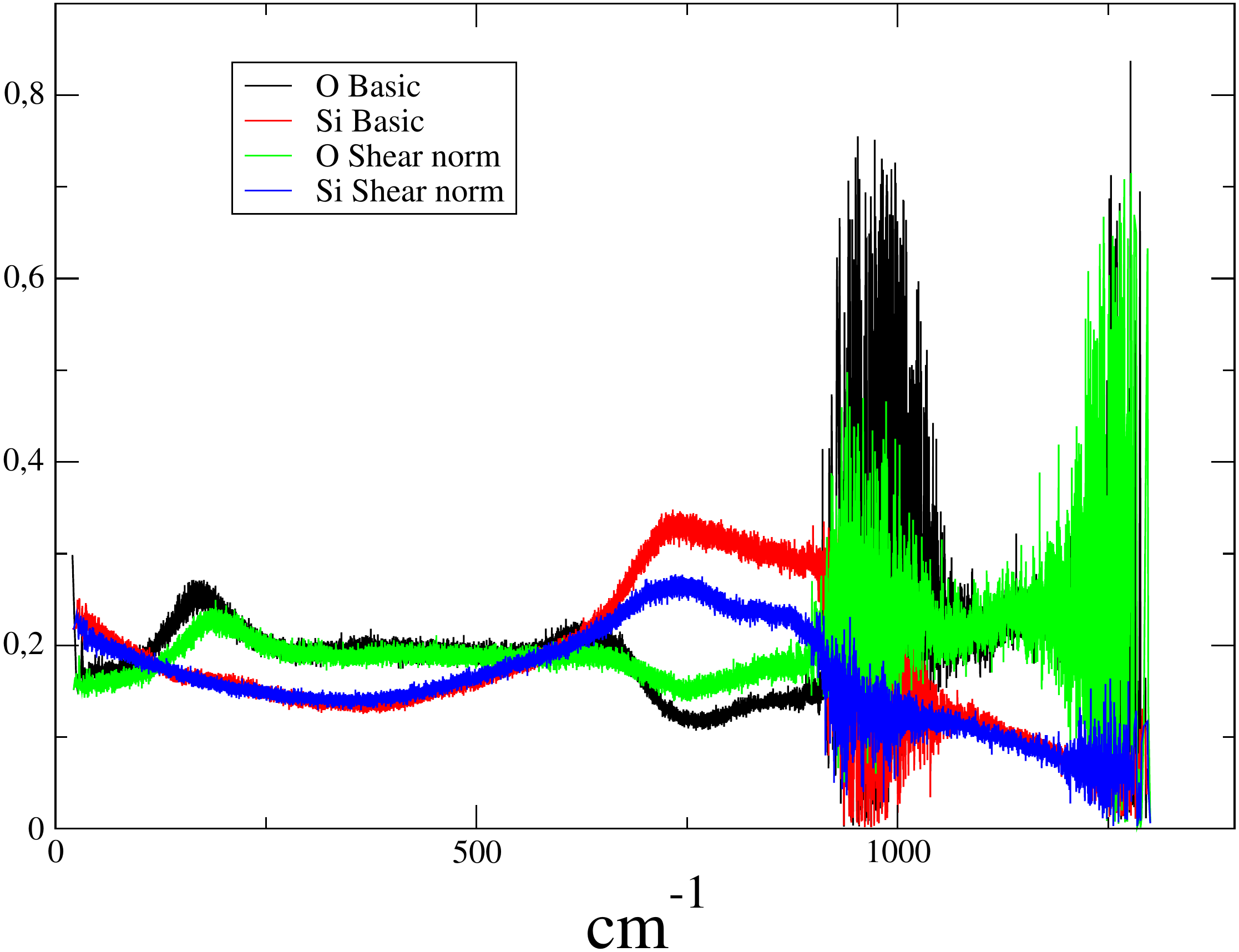}
\caption{ \label{fig-DISP-4-O-Si} Separate contribution of Si \& O atoms in the 4-fold rings, to the displacement field in the eigenvectors as a function of the mode's eigenfrequency, and effect of shear. The curves corresponding to the sheared state have been normalized by the ratio between the number of 4-fold rings in the sheared state, and the number of 4-fold rings in the initial state.}
\end{center}
\end{figure}

\begin{figure}
\begin{center}
\includegraphics[width=8.5cm]{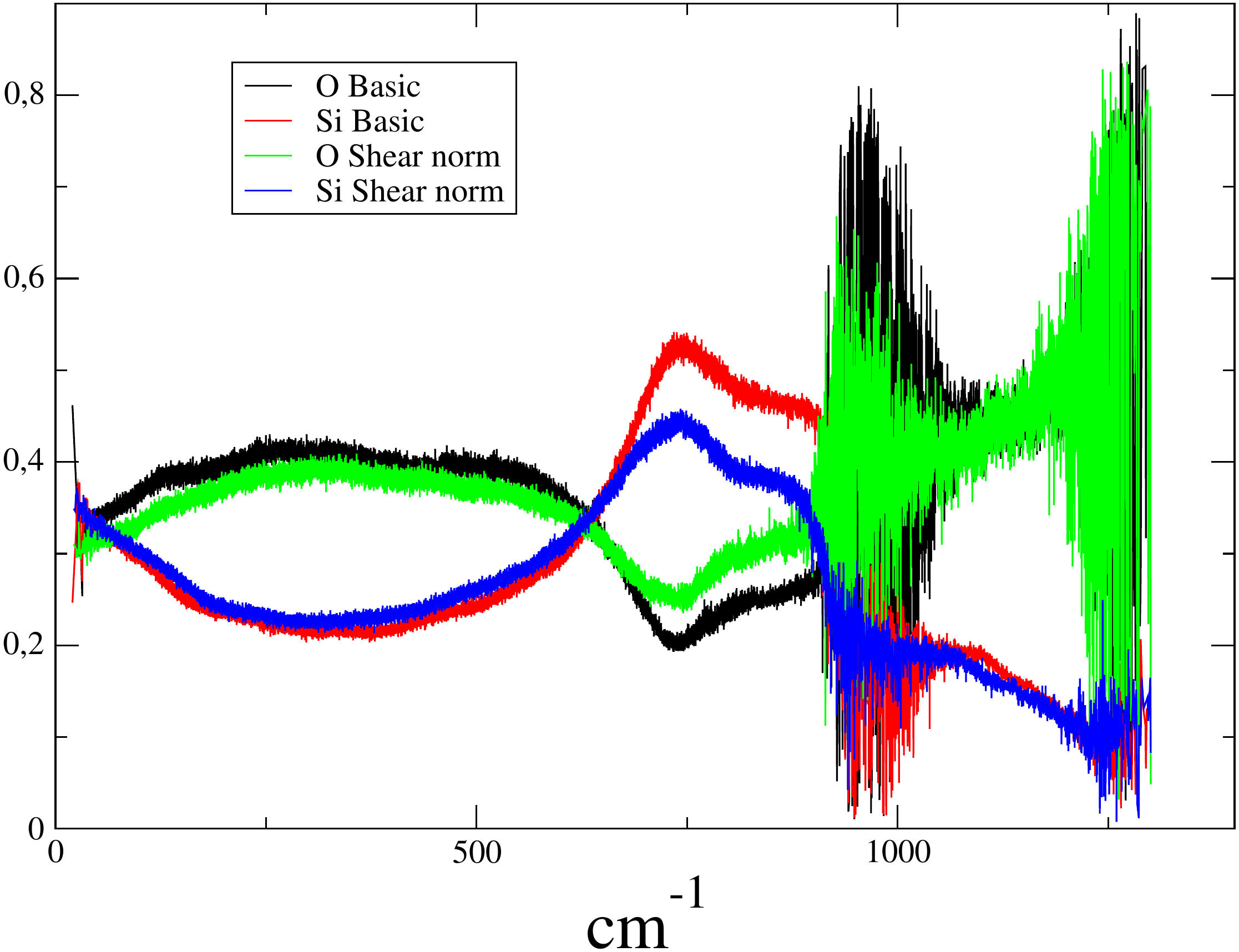}
\caption{ \label{fig-DISP-5-O-Si} Separate contribution of Si \& O atoms in the 5-fold rings, to the displacement field in the eigenvectors as a function of the mode's eigenfrequency, and effect of shear. The curves corresponding to the sheared state have been normalized by the ratio between the number of 5-fold rings in the sheared state, and the number of 5-fold rings in the initial state.}
\end{center}
\end{figure}

\begin{figure}
\begin{center}
\includegraphics[width=8.5cm]{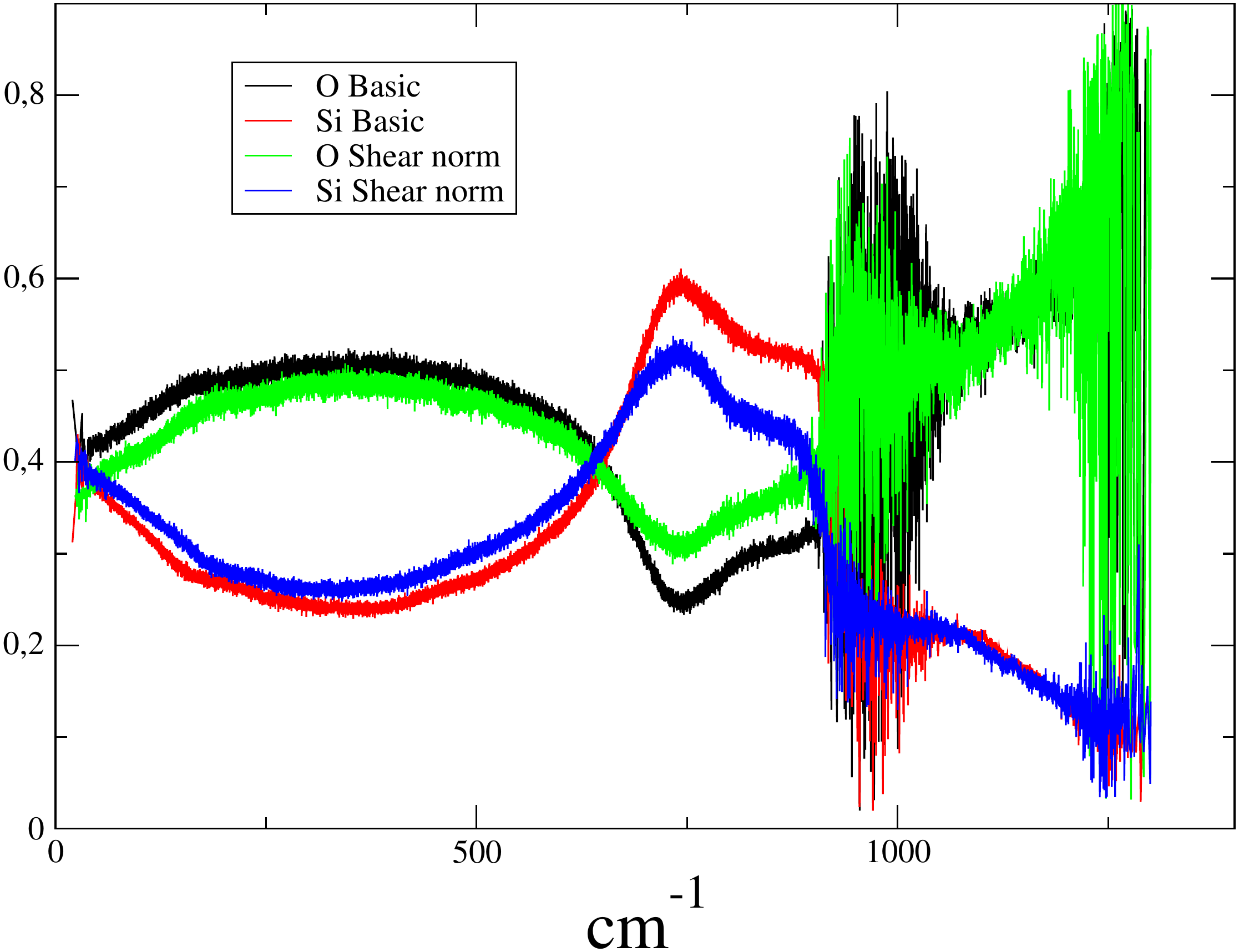}
\caption{ \label{fig-DISP-6-O-Si} Separate contribution of Si \& O atoms in the 6-fold rings, to the displacement field in the eigenvectors as a function of the mode's eigenfrequency, and effect of shear. The curves corresponding to the sheared state have been normalized by the ratio between the number of 6-fold rings in the sheared state, and the number of 6-fold rings in the initial state.}
\end{center}
\end{figure}

In order to describe more precisely the vibration modes in the different characteristic parts of the spectrum, we have decomposed the vibration modes into different contributions: the displacements supported by the different types of atoms (O and Si, Fig.~\ref{fig-DISP-O-Si}), the displacements supported by different kinds of rings (3-fold or 4-fold rings in Fig.~\ref{fig-DISP-3-4}), and the contribution of the vibrations of the different atoms O and Si in the different rings (Figs.~\ref{fig-DISP-3-O-Si} to~\ref{fig-DISP-6-O-Si}). In Fig.~\ref{fig-DISP-O-Si}, we show the amplitude of the displacements supported by the different types of atoms in each vibration modes as a function of the frequency. We see clearly in this figure, that the contributions of the different atoms to the vibrations change mainly in the 700-900~cm$^{-1}$ range, upon plastic shear. Moreover, it is seen in this figure, that there is a transfer from the displacements initially supported by Si atoms that freeze, to displacements of O atoms that increase. 

We will now look at the contribution of the vibrations in the different Si-rings. In the usual interpretations of Raman spectra \cite{alf1998,galeener84,galeener93,galeener85}, D1 band is related to vibrations of 4-fold rings, and D2 band to the breathing mode of 3-fold rings supported by Oxygen atoms, assuming that Si atoms are fixed. However, it should be noted, that to assign definitely the signatures of 3-fold and 4-fold rings is still questionable \cite{smirnov2008}. This was initially related to calculation of eigenfrequencies of isolated rings~\cite{galeener84,galeener93}, and then obtained with ab initio calculations performed on small systems \cite{alf1998}. In Fig.~\ref{fig-DISP-3-4}, we see that the vibrations of 4-fold rings support approximately half of the total displacement, and that 3-fold and 4-fold rings contribute to the displacement field in all the modes and at all frequencies. This can be simply understood by the fact that, in amorphous materials, the vibration modes are in general collective modes, due to the mechanical homogeneity that takes place already at the nanometer scale~\cite{tsamados2009}. Even modes with low mean-free path are referred to as "quasi-local vibration modes" consisting in the superposition of rotational waves. Despite the case of isolated "soft modes" occurring at very low frequencies~\cite{tanguy2010}, it is in general very difficult to relate a vibration mode to the vibration of an isolated entity. Note that the large fluctuations seen in Fig.~\ref{fig-DISP-3-4} at frequencies close to 950 cm$^{-1}$ and 1250 cm$^{-1}$ are due to the very poor statistics in that frequency range were the number of eigenmodes is very low, and to the localization of the vibrations that can - or not - be close to a n-fold ring. In Fig.~\ref{fig-DISP-3-4}, the global increase of the curves upon shear is due to the increase in the number of the small-size rings. Indeed, the ratio between the contribution of the displacements supported by atoms in 3-fold and in 4-fold rings is given mainly by the ratio between the different number of rings (and thus the different number of atoms taken into account when computing the partial displacements amplitude). A rapid comparison between Fig.~\ref{fig-DISP-3-O-Si} and Fig.~\ref{fig-DISP-4-O-Si} shows that the total displacements supported by silicon atoms for example (normalized by the number of rings) is very similar for both kinds of rings: it shows a maximum variation upon shear in the 700-1000 cm$^{-1}$ frequency range. When looking to Fig.~\ref{fig-DISP-3-4}, it is a not surprising to see that 3-fold rings have a maximum contribution at 800 cm$^{-1}$ corresponding to the calculated D2 frequency, while two peaks appear for the 4-fold rings: a peak at 175 cm$^{-1}$ and a peak at 700 cm$^{-1}$. This last peak should correspond to the theoretical D1 band. When looking more precisely to the contribution of the different atoms to the vibrations, we see a finer structure for oxygen atoms. In 4-fold rings (Fig.~\ref{fig-DISP-4-O-Si}), the contribution to the overall displacement of the oxygen atoms shows a first peak at 175 cm$^{-1}$ (explaining the first peak of 4-fold rings displacements) and a secondary peak at 630~cm$^{-1}$, close to the theoretical D1 band, in agreement with the previous analysis (Fig.~\ref{fig-DOS-4f-O-Si}) of the partial density of states. This finer description of the vibrations has a signature in the Raman spectrum (Fig.~\ref{figHHshear}) at the corresponding frequencies.

In general, the curves showing the contribution of the different rings to the displacement field of the eigenmodes flatten upon plastic shear. This does not mean that nothing happens inside the sample. The Figures~\ref{fig-DISP-3-O-Si} to~\ref{fig-DISP-6-O-Si} show the separate contributions of Si and O atoms to the displacement field supported by rings of different sizes, and the corresponding effect of shear. The curves for the sheared configurations have been normalized by the ratio of the number of rings in the sheared vs the number of rings in basic configurations, in order to enhance the differences only due to shear in the shape of these curves, and not due to the quantitative change in the number of units. The common feature in all these curves is the variation occurring in the 700~cm$^{-1}$-1000~cm$^{-1}$ range. In this frequency range, the displacements supported by Si atoms decrease, while the displacements supported by O atoms increase. This means that upon irreversible plastic shear strain, there is an enhancement of the contribution of oxygen motion to the vibrations, while the skeleton of the rings, supported by Si atoms freezes. This is in agreement with the general trend seen in Fig.~\ref{fig-DOS-O-Si}, that shows that, it the same range, the contribution of oxygen atoms to the vibrations increases strongly upon shear, and to the localization of vibrations on oxygen atoms seen in Fig.~\ref{fig-TP-shear}. The effect of the localization at $\sim$850~cm${-1}$ is particularly visible (see Fig.~\ref{fig-DISP-3-O-Si}) in the oxygen motion in 3-fold rings.

\section{DISCUSSION} 

It was shown in this paper that Raman spectra obtained within a semi-classical simulation of large empirical $SiO_2$ samples ($N>$ 8 000 atoms) using the Bond Polarizability model combined with classical calculation of the vibration modes, is able to reproduce the main features of experimental Raman spectra, and shows clearly a sensitivity of the Raman spectra to plastic irreversible shear. The variations of the Raman spectra upon plastic shear are located mainly in the intermediate frequency range (~700-1000~cm$^{-1}$) and in the high frequency range ($>$ 1000~cm$^{-1}$). In the intermediate frequency range, this variation corresponds to enhanced stretching vibrations of oxygen atoms. At smaller frequencies, the global increase of the Raman spectrum can be related to the increase in the number of small rings upon plastic shear. At large frequencies ($>$ 1000~cm$^{-1}$), the contribution of stretching modes supported by oxygen atoms belonging mainly to large order rings decays. 

An additional feature concerns the localization of the breathing vibrations of oxygen atoms on the 3-fold rings at $\sim$800~cm$^{-1}$. This unexpectedly large value for the frequency of the 3-fold breathing modes can be attributed to an excess of rigidity of the Si-O bond in the BKS interactions used~\cite{to2007}. It is however surprisingly enough that the D2 and the 800~cm$^{-1}$ bands (observed experimentally) are well reproduced in our approximate calculation of the Raman spectrum, and even more surprisingly deserve another microscopic description than the usual one~\cite{galeener85}. This remark encourages to perform further studies to check the validity of the bond polarizability model with a systematic comparison of the different empirical descriptions of interatomic interactions, that is far beyond the scope of the present paper.

The computed Raman spectra are sensitive to plastic shear as soon as the sample entered in the plastic flow regime with valuable structural changes. It is not sensitive to the first stages of the plastic deformation as described by plastic instabilities and soft modes~\cite{tanguy2010} whose signature affects probably more specifically low frequency vibration modes measured by Brillouin scattering. The sensitivity to permanent plastic shear may be due to measurable structural changes, especially the increase in the number of low-order rings at a mesoscopic scale. This behaviour is probably composition dependent and it would be interesting to track his signature for other compounds like a-Si, silicate glasses, or even $Ta_2O_5$ glasses known to have a very low mechanical noise~\cite{pinard2011}. Raman spectra are however not sensitive to the amount of plastic deformation, unlike in densification experiments. The difference between shear and densification, is that in the shear case, the nature of defects reaches a stationary states, and second the average stress applied does not vary anymore upon plastic flow. Raman spectra can thus be seen as stress-sensitive sensors.

An interesting contribution of semi-classical models, is that they allow to connect Raman spectra to a complete and detailed description of the large scale vibrational modes. This was done explicitly in our paper. Further work should concern the application of more accurate schemes for assessing vibrational modes. The  accuracy of them should be first establish through comparison with experimental data for systems whose structure is well known~\cite{to2007}. This will open the way to very accurate simulation of Raman spectra including position and intensity of D1 and D2 defect lines. Moreover, perspectives to this work includes in situ  measurement of compressive/shear plasticity at the micrometer scale in silica glasses, that are of crucial interest for crack prediction.

\section{Acknowledgements} 

Authors thank C. Martinet, V. Martinez and B. Champagnon for stimulating discussions and sharing the experimental data. This work was supported by the French Research National Agency program ANR MECASIL. P.U. was supported as visitor at ILM by the Labex IMUST and ANR Initiative d'Excellence.

\bibliography{biblio}

\end{document}